\documentclass[aps,pra,twocolumn,groupedaddress,showpacs,floatfix,superscriptaddress,longbibliography]{revtex4-2}
\usepackage[plainpages=false,pdfpagelabels,colorlinks=true,linkcolor=red,urlcolor=blue,citecolor=blue,pdftitle={Title},pdfauthor={Anubhab Sur, Rubem Mondaini},pdfdisplaydoctitle=true,pdfduplex=DuplexFlipLongEdge]{hyperref}
\usepackage{epsfig}
\usepackage{amsmath,amssymb}
\usepackage{physics}
\usepackage{graphicx}
\usepackage[dvipsnames,usenames]{color}
\usepackage[normalem]{ulem}
\usepackage{soul} 
\usepackage{float}
\usepackage{diagbox}
\usepackage{orcidlink}

\newcommand{\tcr}{\textcolor{black}}

\begin{document}
\title{From Bell products to Greenberger-Horne-Zeilinger states: Quantum memories via emergent Hamiltonians}
\author{Anubhab Sur\,\orcidlink{0009-0007-0359-865X}}
\email{asur3@uh.edu}
\affiliation{Department of Physics,
University of Houston, Houston, Texas 77004, USA}
\affiliation{Texas Center for Superconductivity, University of Houston, Houston, Texas 77004, USA}
\author{Qiujiang Guo\,\orcidlink{0000-0003-1093-3405}}
\email{qguo@zju.edu.cn}
\affiliation{School of Physics, ZJU-Hangzhou Global Scientific and Technological Innovation Center, and Zhejiang Key Laboratory of Micro-nano Quantum Chips and Quantum Control, Zhejiang University, Hangzhou 310058, China}
\author{Rubem Mondaini\,\orcidlink{0000-0001-8005-2297}}
\email{rmondaini@uh.edu}
\affiliation{Department of Physics,
University of Houston, Houston, Texas 77004, USA}
\affiliation{Texas Center for Superconductivity, University of Houston, Houston, Texas 77004, USA}

\begin{abstract}
With the advent of exquisite quantum emulators, storing highly entangled many-body states becomes essential. While entanglement typically builds over time when evolving a quantum system initialized in a product state, freezing that information at any given instant requires quenching to a Hamiltonian with the time-evolved state as an eigenstate, a concept we realize via an emergent Hamiltonian framework. While the emergent Hamiltonian is generically nonlocal and may lack a closed form, we show examples where it is exact and local, thereby enabling, in principle, indefinite state storage limited only by experimental imperfections. Unlike other phenomena, such as many-body localization, our method preserves both local and global properties of the quantum state. In some of our examples, we demonstrate that this protocol can be used to store maximally entangled multiqubit states, such as tensor products of Bell states, or fragile, globally distributed entangled states, in the form of Greenberger-Horne-Zeilinger states, which are often challenging to initialize in actual devices.
\end{abstract}

\maketitle

\section{Introduction}
Storing highly entangled quantum states without degradation is an essential challenge for applications in quantum computing and quantum emulation. In this quest, the many-body localization (MBL) phenomenon~\cite{Nandkishore2015, Alet2018, Abanin2019, Sierant2025}, arising from the application of quenched disorder, has often been thought of as one of the best candidates for storing quantum states and thereby for quantum memory applications~\cite{Smith2016, Carlson2020}. Although MBL is relatively robust against small decoherence~\cite{Levi2016, Fischer2016, Luschen2017}, one of its primary caveats is its inability to preserve most global properties of the state. In particular, the preserved memory achieved via MBL is primarily local, such as the preservation of few-body observables~\cite{Schreiber2015, Choi2016, Kohlert2019, Guo2021a, Scherg2021, Guo2021b, Morong2021}. All nonlocal quantum correlations of the system tend to vary with time, and most of the nonlocal memory of the system fades. Indeed, the total entanglement of the system also grows, albeit slowly~\cite{Bardarson2012, Serbyn2013, Nanduri2014}, signaling nontrivial dynamics and gradual change of the state over time~\footnote{Beyond the intrinsic limitation that MBL protects only local information, its stability itself is under active scrutiny. Numerical work reports slow many-body delocalization beyond one spatial dimension and a localization threshold that drifts with system size, consistent with the absence of stable MBL in two dimensions in the thermodynamic limit~\cite{Doggen2020, Li2025}. Even in one dimension, rare thermal inclusions can seed avalanche mechanisms leading to slow but persistent thermalization~\cite{Peacock2023}, and monitored dynamics can destabilize prethermal localization regimes~\cite{Sun_2025}. These considerations further motivate disorder-free approaches that freeze the full many-body state by construction.}.

A fundamentally different and robust approach is provided by quantum error correction (QEC), which can be set as active~\cite{Shor1995}, wherein quantum information is distributed among many qubits and periodically recovered as needed via an error-reversal procedure~\cite{Knill1997} or passive, in which the entangled state one aims to store is the ground state of a stabilizer Hamiltonian~\cite{Dennis2002}. Unlike MBL, QEC can preserve both local and nonlocal correlations. Nonetheless, large-scale experimental implementation of either passive or active schemes remains challenging~\cite{Fowler2012, Terhal2015}.

Alternatively, in the noisy intermediate-scale quantum (NISQ) era of quantum computation, involving hundreds of imperfect qubits, another major challenge arises from the need to rapidly switch back and forth from idle qubits when not in operation to a strongly interacting regime during multiqubit operations~\cite{Silveri2022}. For instance, a designed multiqubit gate can couple qubits, generating a highly entangled state. Switching off such a gate, effectively promoting an identity operation, can be interpreted as a quantum memory. Yet, this quick switch between strong mutual coupling and idling ("doing nothing") must be done efficiently in a way that there are minimal leaks during the idling phase---current platforms have an estimated error via information leakage in this regime of up to 10\%~\cite{Nico-Katz2024}. 
\begin{figure}[!h] 
  \includegraphics[width=0.6\columnwidth]{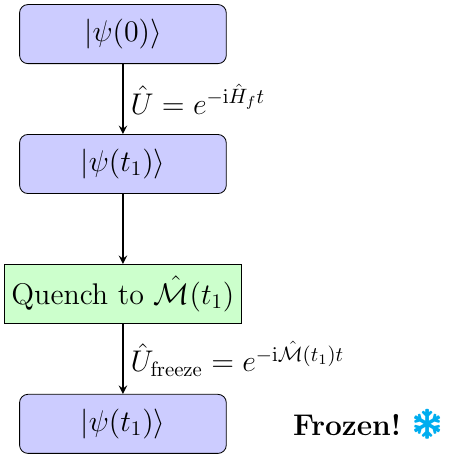}
\caption{Schematics of the protocol. We freeze the dynamics at time $t=t_1$ via carrying out a quench from the entangling Hamiltonian $\hat{H}_f$ to $\hat{\mathcal{M}}(t_1)$---the emergent or freezing Hamiltonian defined at $t_1$.}
\label{fig:fig_1}
\end{figure}

Thus, we seek an alternative approach that can provide stronger and more reliable protection for quantum states, preserving both local and nonlocal correlations more effectively. A mechanism to accomplish that can be obtained by a protocol dubbed "emergent Hamiltonian"~\cite{Vidmar2017,Zhang2021}, wherein after promoting a unitary dynamics generated by an arbitrary Hamiltonian \( \hat{H}_f \) of some generic, often unentangled, initial state, the time-evolved state becomes an eigenstate of an operator \( \hat{\mathcal{M}}(t) \). This operator assumes the form
\begin{equation}
    \hat{\mathcal{M}}(t) = e^{-{\rm i} \hat{H}_f t} \hat{H}_0 e^{{\rm i} \hat{H}_f t} = \hat{H}_0 + \sum_{n=1}^\infty \frac{(-{\rm i} t)^n}{n!} \hat{\mathcal{H}}_n \ ,
    \label{eq:emergent_Hamiltonian}
\end{equation}  
where \( \hat{\mathcal{H}}_n \equiv [\hat{H}_f, [\hat{H}_f, \ldots, [\hat{H}_f, \hat{H}_0] \cdots]] \) is a nested $n$th order commutator, and the operator \( \hat{H}_0 \) is chosen such that the initial state is one of its eigenstates. It is easy to see that $|\psi(t)\rangle$ is an eigenstate of $\hat {\mathcal M}$: $e^{-{\rm i}\hat H_f t} \hat H_0 e^{+{\rm i}\hat H_f t} |\psi(t)\rangle = e^{-{\rm i}\hat H_f t} \hat H_0 |\psi(0)\rangle = E_0 e^{-{\rm i}\hat H_f t} |\psi(0)\rangle = E_0 |\psi(t)\rangle$, where $E_0$ is the energy associated with the initial state in the Hamiltonian $\hat H_0$. However, such a quantity generally leads to a highly non-local operator, precluding its identification as a physical Hamiltonian. Since the series expansion explicitly depends on the time $t$, the short-time dynamics are governed by the first few terms. Typically, the spatial support of the products of the operators involved is extended by each nonvanishing commutator in the series, resulting in the nested commutator being sufficiently local at $t\ll 1$~\cite{Vidmar2017, Zhang2021}, thus giving rise to an emergent physical Hamiltonian only in this regime. 

Building on this, one of our motivations is to develop a protocol capable of pausing the dynamics of a quantum system indefinitely and with perfect precision without necessarily idling the qubits---Fig.~\ref{fig:fig_1} summarizes this. As previously suggested~\cite{Vidmar2017, Zhang2021}, the key idea is that, at any moment during the evolution, the dynamics can be frozen by implementing a quench that transitions from the evolving Hamiltonian \( \hat{H}_f \) to the constructed emergent Hamiltonian \( \hat{\mathcal{M}}(t) \) defined at the exact time  $t$ of the quench, resulting in complete freezing of the evolution, and thus preserving the entire quantum state at that time. 

Although this protocol is universal and, in principle, applicable to freezing \textit{any} unitary dynamics, it is particularly efficient and more meaningful in certain special scenarios, some of which we explore in this article. As mentioned above, the emergent Hamiltonian may be a highly nonlocal operator, which acts as a bottleneck in many cases. Our goal is to explore scenarios where we can derive a compact emergent Hamiltonian exactly at all times, thereby rendering this protocol perfect and exact, as well as others where obtaining a closed-form expression is challenging.

The presentation is divided as follows. In Sec.~\ref{sec:1d}, we introduce the emergent Hamiltonian in a one-dimensional (1D) system and investigate the dynamics of the entanglement entropy in many-body systems. Section \ref{sec:2d} generalizes the exact Emergent Hamiltonians for single-particle systems in higher dimensions, contrasting with an approximate case in the many-body regime. A spectral analysis is carried out in Sec.~\ref{sec:spectral}. We look at preparation and storage of Greenberger-Horne-Zeilinger (GHZ) states in Sec.~\ref{sec:oat_dicke}, and our findings are summarized in Sec.~\ref{sec:conclusion}. Throughout this work we set $\hbar = 1$, and all times are therefore expressed in inverse energy units.

\section{Model in one dimension} \label{sec:1d}
We start with a model Hamiltonian defined on a chain with open boundary conditions,
\begin{equation}
    \hat H_f = \sum_l \left(J_l \hat a_{l+1}^\dagger \hat a_{l}^{} + {\rm H.c.}\right)\ ,
    \label{eq:H_hcb}
\end{equation}
where $\hat a_l^\dagger$ ($\hat a_l^{}$) is the hard-core bosonic creation (annihilation) operator at site $l$. Unlike common bosons, these obey the hard-core constraint $(\hat a_l^{})^2 = (\hat a_l^\dagger)^2=0$, preventing multiple occupancy at a single site~\footnote{Additionally, they exhibit mixed commutation relations, commuting at different sites, $[\hat a_i^{},\hat a_j^{\dagger}]=0$, while anticommuting on the same site, $\{\hat a_i^{},\hat a_i^\dagger\}=1$}. Using the standard mapping between hard-core boson to spin$-1/2$ ladder operators $\hat a_l^\dagger = \hat \sigma_l^+\ (\hat a_l^{} = \hat \sigma_l^-)$~\cite{Matsubara1956} in Eq.~\eqref{eq:H_hcb}, one can associate a lattice with $L$ sites with that of an $L$-qubit system, where each site represents a qubit that is either excited (1) or not (0); the corresponding excitations can hop between consecutive qubits with amplitude $J_l$.

In particular, choosing the hopping amplitudes as $J_l = \frac{1}{2}\sqrt{l(L-l)}$~\cite{Christandl2004, Albanese2004} makes the hopping amplitude matrix $[J_{k,l}]$ resemble that of a large-spin matrix, rendering the Hamiltonian as
\begin{align}
    \hat H_f &= \sum_l\left(\frac{\sqrt{l(L-l)}}{2}\hat \sigma_{l+1}^+ \hat \sigma_l^{-} + {\rm H.c.}\right)\notag \\ &= \frac{\hat S_+ + \hat S_-}{2} =  \hat S_x \ .
\end{align}

Here, $\hat S_\pm$ and $\hat S_x$ are spin operators with spin quantum number $s = \frac{L-1}{2}$, and the second equality holds, in principle, when restricting to the single-excitation subspace. In this one-dimensional case, however, this mapping is still valid for any number of excitations due to the Jordan-Wigner transformation, $\hat a_l = e^{{\rm i}\pi\left(\sum_{m<l}\hat c_m^\dagger\hat c_m^{}\right)}\hat c_l$ with $\hat c_l$ the fermionic annihilation operator, where the system can then be cast as a collection of noninteracting fermions, each of which is mapped by a large spin. That is, using the single-excitation matrix and imposing fermionic statistics via the Slater determinant on single-fermion dynamics, the system effectively describes multiple noninteracting large spins precessing with the same Hamiltonian $\hat H_f = \hat S_x = \sum_l (\frac{\sqrt{l(L-l)}}{2}\hat c_{l+1}^\dagger\hat c_l^{} + {\rm H.c.})$. This is a key distinction from the higher-dimensional case discussed later, where the Jordan-Wigner transformation introduces nontrivial phase factors, preventing a straightforward reduction to a single-particle picture. In two dimensions, therefore, we must rely on the hard-core bosonic description and consider the full many-body Hamiltonian.

\begin{figure*}[!tb] 
  \includegraphics[width=0.9\textwidth]{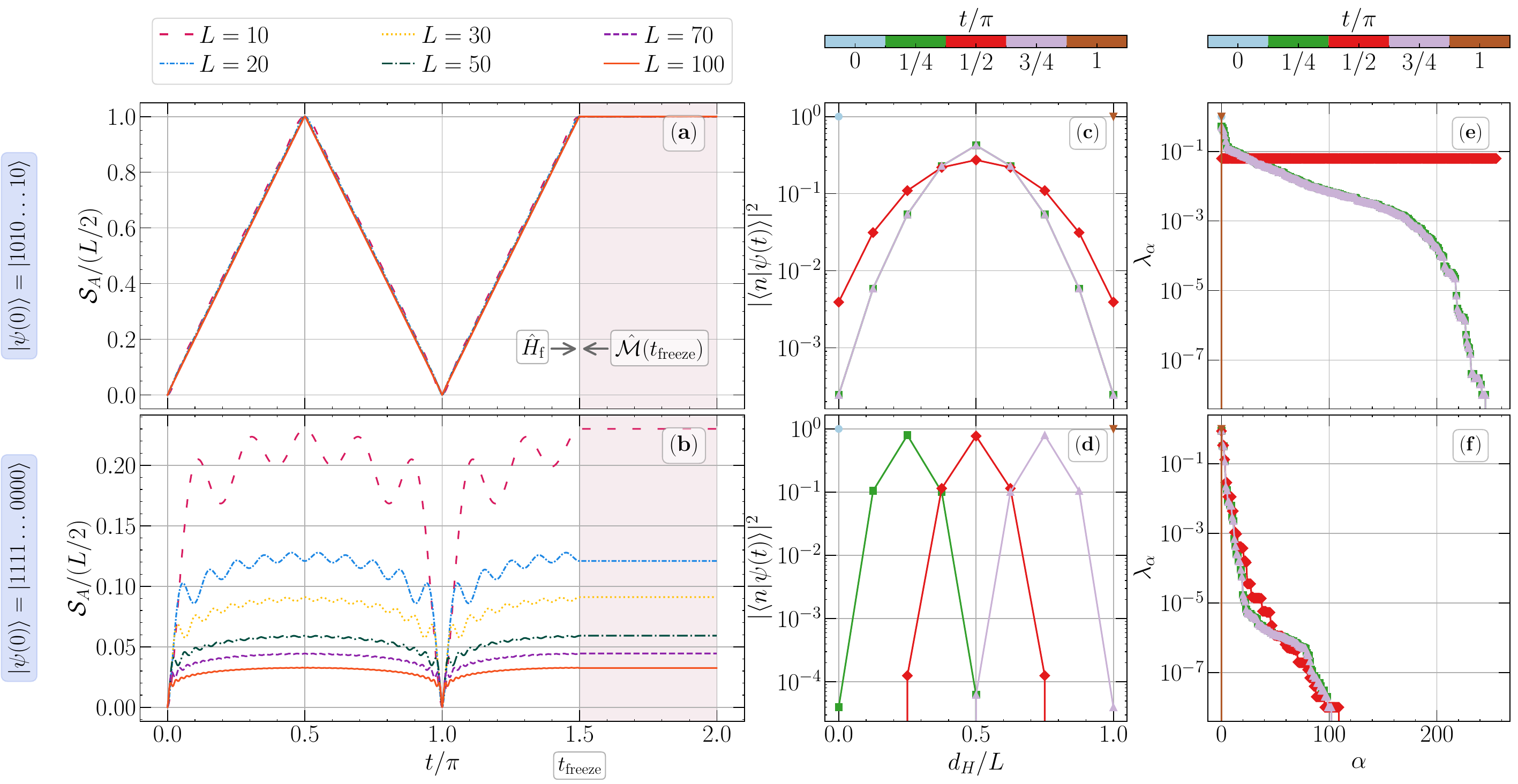}
  \caption{Dynamics of the half-chain entanglement entropy, scaled by the maximal entropy, considering different chain sizes, for (a) a density-wave [(b) domain wall] initial product state $|\psi(0)\rangle$; a quench $\hat H_{f}\to\hat {\cal M}$ is performed at $t_{\rm freeze}=3\pi/2$. (c) and (d) Distribution of Hamming distances and the corresponding weights $|\langle n|\psi(t)\rangle|^2$ in the time-dependent state $|\psi(t)\rangle$ of the Fock states $|n\rangle$, at representative times in the dynamics. (e) and (f) Schmidt coefficients of the half-chain decomposition at similar times. In (c) and (d), the system size is $L=16$; as indicated, (a), (c), and (e) [(b), (d), and (f)] refer to the density-wave [domain wall] initial product state.}
    \label{fig:fig2}
\end{figure*}

Following the prescription in Ref.~\cite{Vidmar2017}, we calculate the emergent Hamiltonian $\hat {\mathcal M}(t)$, taking the initial state as an eigenstate of a simple, diagonal in the computational basis, Hamiltonian $\hat{H}_0$,
\begin{equation}
    \hat{H}_0 = \sum_{l = 0}^{L-1} l\ \hat a_l^{\dagger}\hat a_l^{\phantom{\dagger}}= \sum_{l = 0}^{L-1} l\ \hat{c}_{l}^{\dagger}\hat c_l^{\phantom{\dagger}}= -\hat S_z+\frac{L-1}{2}\hat N\ ,
\end{equation}
where we used the Jordan-Wigner transformation in the second equality, and $\hat N = \sum_l \hat{c}_{l}^{\dagger}\hat c_l^{\phantom{\dagger}}$ is the total particle number operator. Unlike the case of homogeneous hoppings~\cite{Zhang2021}, in this case, the series that results in $\hat {\mathcal M}(t)$ [Eq.~\eqref{eq:emergent_Hamiltonian}] can be precisely summed by making the identifications of the nested commutators, $\hat {\cal H}_{2n} = -\hat S_z$ and $\hat {\cal H}_{2n+1} = {\rm i}\hat S_y$, when using the standard spin commutation relations, $[\hat S_\alpha,\hat S_\beta]={\rm i}\epsilon_{\alpha\beta\gamma}\hat S_\gamma$, yielding
\begin{align}
\hat{\mathcal{M}}(t) 
  &= -\hat S_z + \frac{L-1}{2} \hat N\nonumber \\
  &\quad + \sum_{n=1,3,\ldots}^{\infty} \frac{(-{\rm i} t)^n}{n!}\, {\rm i} \hat S_y 
       - \sum_{n=2,4,\ldots}^{\infty} \frac{(-{\rm i} t)^n}{n!}\, \hat S_z \nonumber \\
  &= \frac{L-1}{2}\hat N + \sin(t)\, \hat S_y - \cos(t)\, \hat S_z \, .
  \label{eq:emergent_1d}
\end{align}

Thus, we provide an analytically exact expression of the emergent many-body Hamiltonian in a quadratic, local form---a sum of diagonal and nearest-neighbor "hopping" terms in Eq.~\eqref{eq:emergent_1d}, when casting in terms of hard-core bosonic operators:
\begin{eqnarray}
\hat{\mathcal{M}}(t) &=&\sum_{l=0}^{L-1} \left[ \frac{L-1}{2} + \cos(t)\left( l - \frac{L-1}{2} \right) \right] \hat{a}_l^\dagger \hat{a}_l^{\phantom{\dagger}} \nonumber \\
  &\quad&+ \sin(t) \sum_{l=1}^{L-1} \left( \frac{{\rm i} \sqrt{l(L-l)}}{2} \hat{a}_l^\dagger \hat{a}_{l-1}^{\phantom{\dagger}} + {\rm H.c.}\right).
\end{eqnarray}

A fundamental characteristic of the dynamics generated by $\hat H_f$ is that it enables the preparation of highly entangled states from simple, easily initialized separable states such as product states. For example, Fig.~\ref{fig:fig2} illustrates the time evolution of the half-chain entanglement entropy ${\cal S}_A$ [${\cal S}_A = -\sum_\alpha \lambda_\alpha^2\log_2 \lambda_\alpha^2$, where $\lambda_\alpha$ are Schmidt coefficients] for two initial product states~\footnote{We compute the entanglement entropy via the correlation matrix of free fermions for the relevant partition, as illustrated in Ref.~\cite{Peschel2009}}, a density wave, $|\psi(0)\rangle = |1010\ldots10\rangle$ [Fig.~\ref{fig:fig2}(a)] and a domain wall $|\psi(0)\rangle = |111\ldots000\rangle$ [Fig.~\ref{fig:fig2}(b)]  --- any product state is an eigenstate of $\hat H_0$. Starting from these separable initial states, ${{\cal S}_A(t=0)=0}$, the entanglement entropy and $|\psi(t)\rangle$ evolve periodically with a fundamental period of $2\pi$, consistent with expectations from the spin mapping. At odd multiples of $\pi$, the roles of occupied and unoccupied sites are interchanged, yielding another separable state, i.e., $|\psi\left(t=(2n+1)\pi\right)\rangle = \prod_l \hat \sigma^{(l)}_x |\psi(t=0)\rangle$, where $n \in \mathbb{Z}_{\geq 0}$. This can be confirmed by the computation of the Hamming distance between these two product states, $d_{H}(|n\rangle,|m\rangle) \equiv \sum_{l=0}^{L-1}|n_l-m_l|$ with $|n\rangle = |n_0\ n_1\ldots n_{L-1}\rangle$ and $|m\rangle = |m_0\ m_1\ldots m_{L-1}\rangle$, showing that $d_H(|\psi(t=0)\rangle,|\psi(t = \pi)\rangle ) = L$ [Figs.~\ref{fig:fig2}(c) and \ref{fig:fig2}(d)] --- it relates to the perfect quantum state transfer protocol~\cite{Albanese2004}, now applied to a many-body state.

Notably, for the case of the density-wave initial state at times odd multiples of \( \pi/2 \), ${\cal S}_A$ reaches its peak [Fig.~\ref{fig:fig2}(a)], corresponding to maximally entangled states with the highest possible Schmidt rank $r = 2^{L/2}$, with all Schmidt coefficients $\lambda_\alpha$ in the half-chain decomposition $|\psi(t)\rangle = \sum_{\alpha=0}^r\lambda_\alpha|n_A\rangle\otimes|n_B\rangle$ being equal [Fig.~\ref{fig:fig2}(e)]---a property never observed for domain wall initial states [Fig.~\ref{fig:fig2}(f)], explaining why the entanglement entropy is typically small for the latter. Thus, by quenching the dynamics at special times, $t_{\rm freeze} = (2n+1)\pi/2$, from $\hat H_f$ to $\hat {\mathcal M}(t_{\rm freeze})$ in Eq.~\eqref{eq:emergent_1d}, one can efficiently prepare and store maximally entangled states when starting from a simple density-wave product state, as exemplified in Fig.~\ref{fig:fig2}(a). 

In particular, the nature of these states can be compactly written (for an even $L$) as 
\begin{align}
|\psi(t=\pi/2)\rangle &=
\bigotimes_{l=0}^{\tfrac{L}{2}-1} 
\frac{|1\rangle_{l}\,|0\rangle_{L-1-l}-{\rm i}\,|0\rangle_{l}\,|1\rangle_{L-1-l}}{\sqrt{2}}
\notag \\ &=\bigotimes_{l=0}^{\tfrac{L}{2}-1} |\Psi\rangle_{l,\,L-1-l} \ ,
\end{align}
namely, a tensor product of pairwise entangled states $|\Psi\rangle \equiv \tfrac{|10\rangle-{\rm i}|01\rangle}{\sqrt{2}}$ between inversion-symmetric qubits in the chain. A similar state with a different phase is obtained at time $t=3\pi/2$.

From an experimental perspective, preparing such a nonlocal pairwise entangled state is straightforward only with all-to-all connectivity, where each pair can be generated in parallel through standard initialization of Bell-like states~\footnote{That is, starting from $|00\rangle$ through a Hadamard gate on the first qubit, a controlled-\textsc{not} on the pair, an \emph{X} gate on the second qubit, and a subsequent phase gate $S$ on the second qubit~\cite{Nielsen_Chuang_2010}}, requiring a circuit depth of four layers. The great majority of quantum circuits use a local (often nearest neighbor) connectivity~\cite{Arute2019, Guo2021b, Xiang2024}, however. As such, creating such long-range entanglement necessarily entails a circuit depth that is extensive in the number of qubits. Our approach, on the other hand, even leveraging only nearest-neighbor connectivity, enables the creation of this state via unitary dynamics generated by $\hat H_f$ and its subsequent storage by quenching to the corresponding emergent Hamiltonian $\hat{\mathcal M}(t)$ [Eq.~\eqref{eq:emergent_1d}].

\section{Two Dimensions} \label{sec:2d}

We now extend the entangling Hamiltonian $\hat H_f$ to a two-dimensional (2D) lattice with $L_x \times L_y$ sites:

\begin{equation}\label{eq:Hf_2dNN}
\begin{aligned}
\hat H_f
  ={}& \sum_{\mathbf l}\frac{\sqrt{l_x(L_x-l_x)}}{2}\!\left(\hat a_{{\mathbf l}+\hat x}^{\dagger}\hat a_{\mathbf l}^{\phantom{\dagger}} + \mathrm{H.c.}\right) \\
  &+ \sum_{\mathbf l}\frac{\sqrt{l_y(L_y-l_y)}}{2}\!\left(\hat a_{{\mathbf l}+\hat y}^{\dagger}\hat a_{\mathbf l}^{\phantom{\dagger}} + \mathrm{H.c.}\right) \\
  ={}& \hat S_{1x} \otimes \hat I_2 + \hat I_1 \otimes \hat S_{2x},
\end{aligned}
\end{equation}

\begin{table}[t]
\caption{Comparison of approaches to extract the emergent Hamiltonian for single and many excitations in 1D and 2D systems.}
\label{table:approaches}
\begin{ruledtabular}
\begin{tabular}{lcc}
 & \textbf{Single excitation} & \textbf{Many particles} \\
\colrule
\textbf{1D} & Exact closed form   & Exact closed form   \\
\textbf{2D} & Exact closed form   & Approximate methods \\
\end{tabular}
\end{ruledtabular}
\end{table}
\noindent where ${\bf l}=(l_x,l_y)$, and the summation takes into account the open boundary conditions. As previously advanced, the second equality holds only when we restrict to the single-particle subspace, since the application of the Jordan-Wigner transformation in two dimensions introduces nontrivial phase factors that significantly complicate the representation, ruling out this compact form in the many-body regime. Consequently, a simple mapping to fermionic operators is unattainable, which obstructs a direct extrapolation from single-particle behavior to many-body dynamics via Slater determinants. 
As such, it precludes the possibility of obtaining an exact closed form in the many-body regime in dimensions larger than one---Table \ref{table:approaches} summarizes the possible scenarios.

In what follows, to better understand the interplay between dimensionality and interaction effects, we separately analyze the single- and the many-particle cases. For the latter, we rely on approximations and investigate their validity.

\subsection{Single-particle case}

Following the same reasoning as in one dimension, we assume that the original Hamiltonian \( \hat H_0 \), one of whose eigenstates serves as the initial state for the dynamics, is given by:

\begin{align}
    \hat H_0 =& \sum_{\bf l} (l_x + l_y)\hat a_{\bf l}^\dagger \hat a_{\bf l}^{\phantom{\dagger}} \notag \\
    =& -\hat S_{1z} \otimes \hat I_2 -\hat I_1\otimes \hat S_{2z}+\left(\frac{L_x+L_y}{2}-1\right)\hat I_1\otimes\hat I_2\ .
    \label{eq:H_0_2d}
\end{align}
Repeating the previous calculations, we obtain the following emergent Hamiltonian,
\begin{align}
\hat{\mathcal{M}}(t) &= 
\left(\tfrac{L_x+L_y}{2}-1\right)\hat I_1 \otimes \hat I_2 \notag \\
&\quad + \sin(t)\!\left(\hat S_{1y}\otimes \hat I_2 
   + \hat I_1 \otimes \hat S_{2y}\right) \notag \\
&\quad - \cos(t)\!\left(\hat S_{1z}\otimes \hat I_2 
   + \hat I_1 \otimes \hat S_{2z}\right)\  ,
   \label{eq:M_NN}
\end{align}
where the nested commutators used to compute the infinite series [Eq.~\eqref{eq:emergent_Hamiltonian}] follow the pattern $\hat {\cal H}_{2n} = -(\hat S_{1z}\otimes\hat I_2 + \hat I_1\otimes\hat S_{2z})$ and $\hat {\cal H}_{2n+1} ={\rm i}(\hat S_{1y}\otimes\hat I_2 + \hat I_1\otimes\hat S_{2y})$, for even and odd terms, respectively. This result establishes a direct analogy between the single-particle dynamics of a hard-core boson and that of two noninteracting large spins with spin quantum numbers \( s_1 = \frac{(L_x-1)}{2} \) and \( s_2 = \frac{(L_y-1)}{2} \), respectively. Here, each direction in the original model maps an independent spin, and is a direct generalization of the emergent Hamiltonian for the single-particle version of the 1D case, Eq.~\eqref{eq:emergent_1d}.

In fact, interactions between these spins can be introduced by incorporating diagonal hopping terms in the original Hamiltonian, by modifying \( \hat H_f \) as
\begin{align}
\hat H_f  =  \hat S_{1x} \otimes \hat I_2 + \hat I_1 \otimes \hat S_{2x} + \hat S_{1x} \otimes \hat S_{2x}\ .
\label{eq:Hf_2dNNN}
\end{align}

\begin{figure*}[!tb] 
  \includegraphics[width=0.9\textwidth]{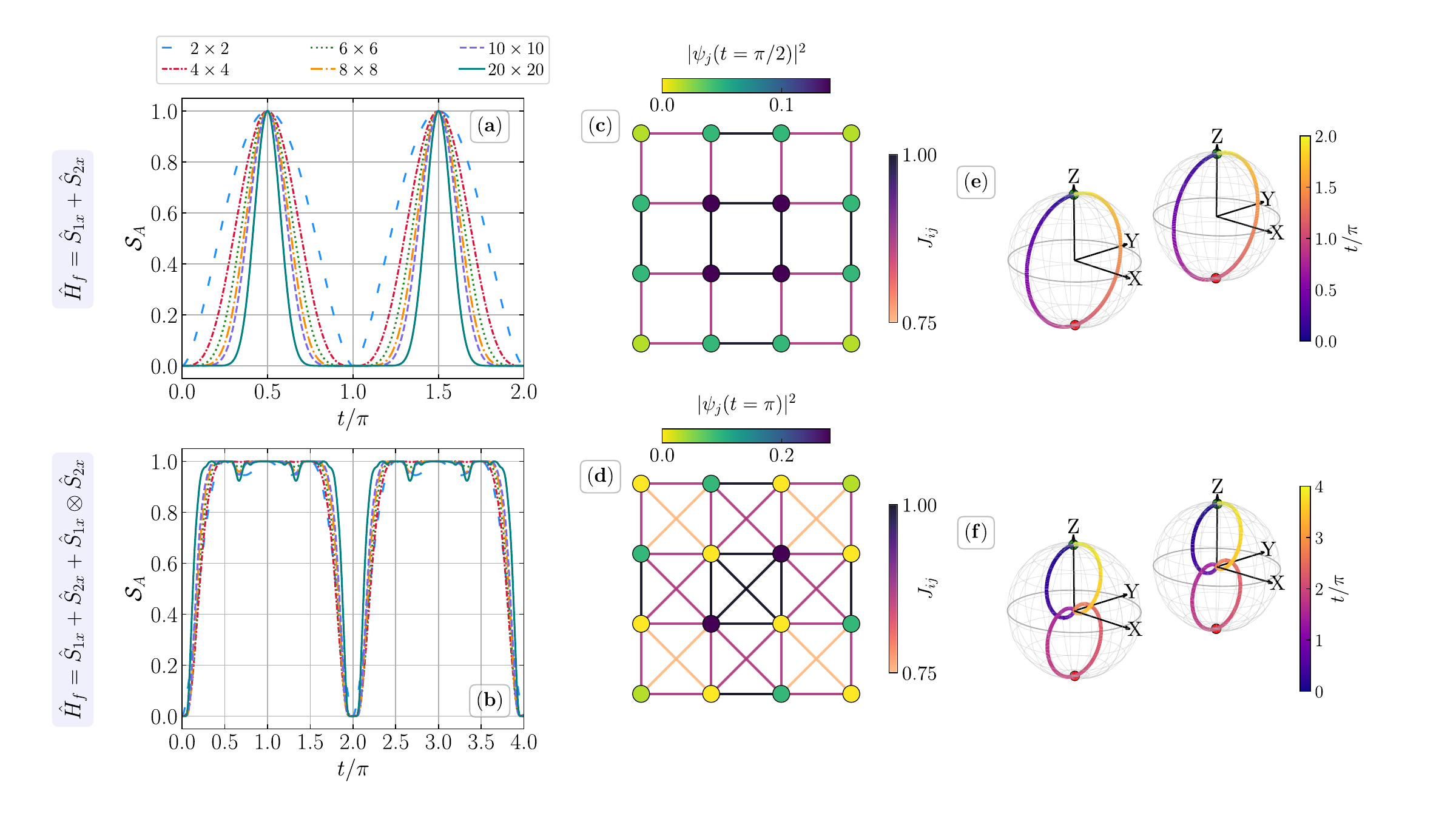}
  \caption{Dynamics of the half-system entanglement entropy considering different lattice sizes $L_x\times L_y$, for the [next-] nearest-neighbor entangling Hamiltonian (a) [(b)] taking as initial product state $|\psi(0)\rangle $ a single excitation at the lower left corner of the lattice. (c) and (d) Snapshots of the site resolved probabilities $|\psi_j|^2$ at the times where the entanglement entropy is maximal. (e) and (f) Expectation values of $\{\langle \hat S_{1\alpha}\rangle\}$ and $\{\langle \hat S_{2\alpha}\rangle\}$ ($\alpha=x,y,z$) over time, defining a trajectory in the Bloch sphere---here, the colorbar maps the time $t$. As indicated, (a), (c), and (e) [(b), (d), and (f)] refer to the entangling Hamiltonian $\hat H_f$ in Eq.~\eqref{eq:Hf_2dNN} [\eqref{eq:Hf_2dNNN}].}
    \label{fig:fig_3}
\end{figure*}

This compact representation highlights that the nearest-neighbor hopping amplitudes in this case are $J_{l_\mu} = \frac{1}{2}\sqrt{l_\mu(L_\mu-l_\mu)}$ with $\mu = x,y$ in terms of hard-core boson operators, and that the next-nearest-neighbor amplitudes are $J_l = \frac{1}{4}\sqrt{l_xl_y(L_x-l_x)(L_y-l_y)}$. Within this framework, a convergent emergent Hamiltonian can still be obtained, expressed as~\footnote{Here, we simplify the notation by omitting tensor products, which are obvious in the context.},
\begin{align}
\hat{\mathcal{M}}(t) 
 &= \left(\tfrac{L_x+L_y}{2}-1\right) \notag \\
 &\quad + \sin\!\left[(1+\hat S_{1x})t\right] \hat S_{2y} 
   - \cos\!\left[(1+\hat S_{1x})t\right] \hat S_{2z} \notag \\
 &\quad + \hat S_{1y}\,\sin\!\left[(1+\hat S_{2x})t\right] 
   - \hat S_{1z}\,\cos\!\left[(1+\hat S_{2x})t\right] \, ,
\end{align}
where the nested commutators in this case read $\hat {\cal H}_{2n}=-[\hat S_{1z}(1+\hat S_{2x})^{2n}+(1+\hat S_{1x})^{2n}\hat S_{2z}]$  and $\hat {\cal H}_{2n+1}={\rm i} [\hat S_{1y}(1+\hat S_{2x})^{2n+1}+(1+\hat S_{1x})^{2n+1}\hat S_{2y}]$, which, as before, can be exactly summed in Eq.~\eqref{eq:emergent_Hamiltonian}.

Figures \ref{fig:fig_3}(a) and \ref{fig:fig_3}(b) show the evolution of the half-system entanglement entropy ${\cal S}_A$ under the dynamics generated by the Hamiltonian $\hat H_f$ in Eqs.~\eqref{eq:Hf_2dNN} and \eqref{eq:Hf_2dNNN}, respectively, when taking the initial state as a single excitation in one corner of the lattice, $|\psi(0)\rangle = |\ldots001\rangle$. Here, the partition $A$ is drawn to pick the bottom half of the sites in the lattice. At certain times, $t=t_{\rm max \ {\cal S}_A}=\pi/2$ for the `nearest neighbor' $\hat H_f$ [Eq.~\eqref{eq:Hf_2dNN}] and $t_{\rm max \ {\cal S}_A}=\pi$ for the `next-nearest neighbor one [Eq.~\eqref{eq:Hf_2dNNN}] \footnote{The period difference depends only on whether \(L_x\) or \(L_y\) is even: mapping to large spins \(s_i=(L_i-1)/2\) makes \(s_i\) half-integer whenever \(L_i\) is even. In the \(\hat S_{i,x}\) eigenbasis, with \(m_i=-s_i,-s_i+1,\ldots,s_i\) (unit spacing), the spectrum is \(E=m_1+m_2\) in the nearest-neighbor case and \(E=m_1+m_2+m_1m_2\) in the next-nearest case. The recurrence time (ignoring any overall phase) is set by the fundamental spacing of energy gaps \(g\), i.e., the positive number for which every difference \(E-E'\) is an integer multiple of \(g\) (here this equals the smallest nonzero gap). For the nearest-neighbor case the gap set is \(\mathbb{Z}\), so \(g=1\) and the period is \(2\pi\). For the next-nearest case, shifting \(m_1\!\to\! m_1+1\) gives \(\Delta E=1+m_2\). Thus, the gap set is \(\mathbb{Z}\) if both spins are integer (both \(L_x,L_y\) odd), but becomes \(\mathbb{Z}+\tfrac12\) as soon as one spin is half-integer (i.e., at least one of \(L_x,L_y\) is even). So the period is \(2\pi\) in the former (\(g=1\)) and \(4\pi\) in the latter (\(g=\tfrac12\)). At \(4\pi\) the evolution adds a global phase (with respect to the initial state) of \(-1\), so observables already recur, while the unitary returns to identity only at \(8\pi\). For the nearest-neighbor case, it already returns at \(2\pi\)}, the half-system entanglement entropy reaches its maximal value, ${\cal S}_A = 1$ (in bits).

These instants correspond to half the time needed to achieve a quantum state transfer, wherein the excitation is transferred to the opposite corner of the lattice~\cite{Christandl2004, Xiang2024}. In particular, at maximum entanglement, the state has a distribution over many sites [Figs.~\ref{fig:fig_3}(c) and \ref{fig:fig_3}(d)], in a way to obey inversion symmetry implicit in $\hat H_f$, ${\bf l}=(l_x,l_y)\to (L_x-1-l_x,L_y-1-l_y)={\bf l^\prime}$, or that the spin projection quantum number along the $z$ axis $m_{1,2}=-s_{1,2},\ldots,s_{1,2}$ for both spins are simultaneously reversed. In this latter mapping, one can interpret the dynamics as the precession of (the expectation values of) spins about the $x$ axis in the Bloch sphere with the same rate [Figs.~\ref{fig:fig_3}(e) and \ref{fig:fig_3}(f)]. While in the nearest neighbor, or noninteracting spins case $\hat H_f = \hat S_{1x} +\hat S_{2x}$, the precession is trivial, it becomes more involved in the presence of interactions [Eq.~\eqref{eq:Hf_2dNNN}], yet also accomplishing the quantum state transfer protocol when the trajectory, starting at $+z$ for our initial state reaches the $-z$-pole of the Bloch sphere~\cite{Xiang2024}.

Returning to the analysis of the emergent Hamiltonian, it becomes clear that one can utilize it to freeze the dynamics after accomplishing the quantum state transfer. For instance, the original perfect protocol of quantum state transfer relies on switching off all couplings once the transfer is accomplished. Instead, one can quench the dynamics,  $\hat H_f\to \hat {\mathcal M}(2t_{\rm max \ {\cal S}_A})$, since the transferred state is an eigenstate of the emergent Hamiltonian. This is immediately seen for the nearest-neighbor case [Eq.~\eqref{eq:M_NN}], in which $\hat {\mathcal M}(t = \pi) = \left(\tfrac{L_x+L_y}{2}-1\right) +\cos(t)\!\left(\hat S_{1z}+  \hat S_{2z}\right)$, and the transferred state, $|\psi\rangle = |-z_1\rangle\otimes|-z_2\rangle$, is clearly an eigenstate.

\subsection{Many-particle case} \label{sec:many_particle}

We now extend our analysis to the many-particle case on the 2D lattice, where interactions due to the hard-core constraint significantly complicate the dynamics. Unlike in the single-particle scenario, where we obtained an exact closed-form emergent Hamiltonian, the many-body case presents formidable challenges due to the increasing complexity of the nested commutators in Eq.~\eqref{eq:emergent_Hamiltonian} and the impossibility of mapping the Hamiltonian to that of large spins. This results in the compactness of the emergent Hamiltonian being elusive and necessitating the use of approximate methods to gain meaningful insights.

For example, consider the entangling Hamiltonian
\begin{align}
\hat{H}_f &= \sum_{\mathbf l}\frac{\sqrt{l_x\!\left(L_x-l_x\right)}}{2}
 \left(\hat a_{\mathbf l+\hat{x}}^{\dagger}\hat a_{\mathbf l}^{\phantom{\dagger}}+{\rm H.c.}\right) \notag \\
&\quad + \sum_{\mathbf l}\frac{\sqrt{l_y\!\left(L_y-l_y\right)}}{2}
 \left(\hat a_{\mathbf l+\hat{y}}^{\dagger}\hat a_{\mathbf l}^{\phantom{\dagger}}+{\rm H.c.}\right) \notag \\
&\quad + J_{\times}\sum_{\mathbf l}\Bigl[
 \left(\hat a_{\mathbf l-\hat{x}+\hat{y}}^{\dagger}\hat a_{\mathbf l}^{\phantom{\dagger}}+{\rm H.c.}\right) \notag \\
&\qquad\qquad\quad
 + \left(\hat a_{\mathbf l+\hat{x}+\hat{y}}^{\dagger}\hat a_{\mathbf l}^{\phantom{\dagger}}+{\rm H.c.}\right)
 \Bigr]\ ,
 \label{eq:Hf_NNN}
\end{align}
which, unlike previous cases, sets a homogeneous next-nearest-neighbor (NNN) hopping, with amplitude $J_{\times}$. And, as before, the initial Hamiltonian, one of whose eigenstates serves as the starting point for the dynamics, is given by $\hat{H}_0 = \sum_{\bf l}\left(l_x+l_y\right)\hat{a}_{\bf l}^{\dagger}\hat{a}_{\bf l}^{\phantom{\dagger}}$. Summing Eq.~\eqref{eq:emergent_Hamiltonian} does not lead to a closed form, as previously realized in other 1D Hamiltonians~\cite{Vidmar2017, Zhang2021}. Instead, another approach is to truncate that sum at either first, $\hat{\mathcal{M}}^{(1)}(t) = \hat H_0 - \mathrm{i}\,t\,\hat{\mathcal H}_1$, or second orders in time, $\hat{\mathcal{M}}^{(2)}(t) = \hat H_0 - \mathrm{i}\,t\,\hat{\mathcal H}_1 - \frac{t^2}{2}\,\hat{\mathcal H}_2$. For sufficiently small times, $t$ can be treated as a perturbative parameter, and truncating at these orders should provide reasonable approximations within short-time dynamics. Notably, with each higher order $n$, $\hat{\mathcal{M}}^{(n)}(t)$ becomes increasingly nonlocal, spreading across the lattice and appearing structurally complex. The explicit forms of the $\hat{\mathcal{M}}^{(1)}(t)$ and $\hat{\mathcal{M}}^{(2)}(t)$ are provided in Appendix \ref{sec:appA}, where we anticipate that they involve current-density-like terms for $\hat{\mathcal H}_1$ and density-assisted hoppings for $\hat{\mathcal H}_2$, supplemented by a renormalization of the onsite energies.

\begin{figure*}[t!] 
  \includegraphics[width=0.9\textwidth]{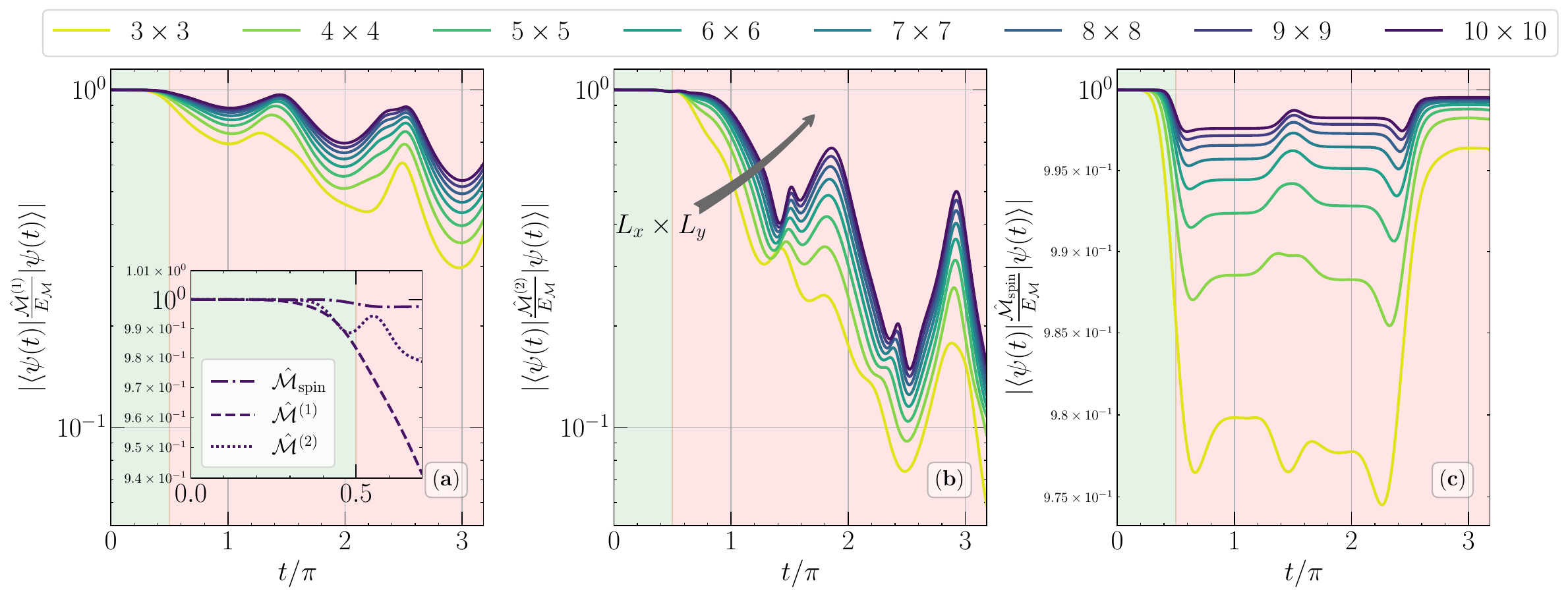}
\caption{(a) Time evolution of the overlap between the time-evolved state \(|\psi(t)\rangle\) and the normalized \(\hat{\mathcal{M}}(t)|\psi(t)\rangle\) under the first-order emergent Hamiltonian $\hat{\mathcal{M}}^{(1)}$ for different system sizes $L_x\times L_y$. The initial state consists of two particles positioned at opposite corners of the square lattice. (b) [(c)] The same, but considering the second-order emergent Hamiltonian $\hat{\mathcal{M}}^{(2)}$ [the spin emergent Hamiltonian \(\hat{\mathcal{M}}_{\text{spin}}\)]. The inset in (a) contrasts the overlap at short times among the different approximate emergent Hamiltonian for a $10\times 10$ lattice. The shadings in all panels highlight the different regimes where the dynamics are effectively single particle and when such a distinction can no longer be made---see text.}
\label{fig:fig_4}
\end{figure*}

\begin{figure}[!b] 
  \includegraphics[width=\columnwidth]{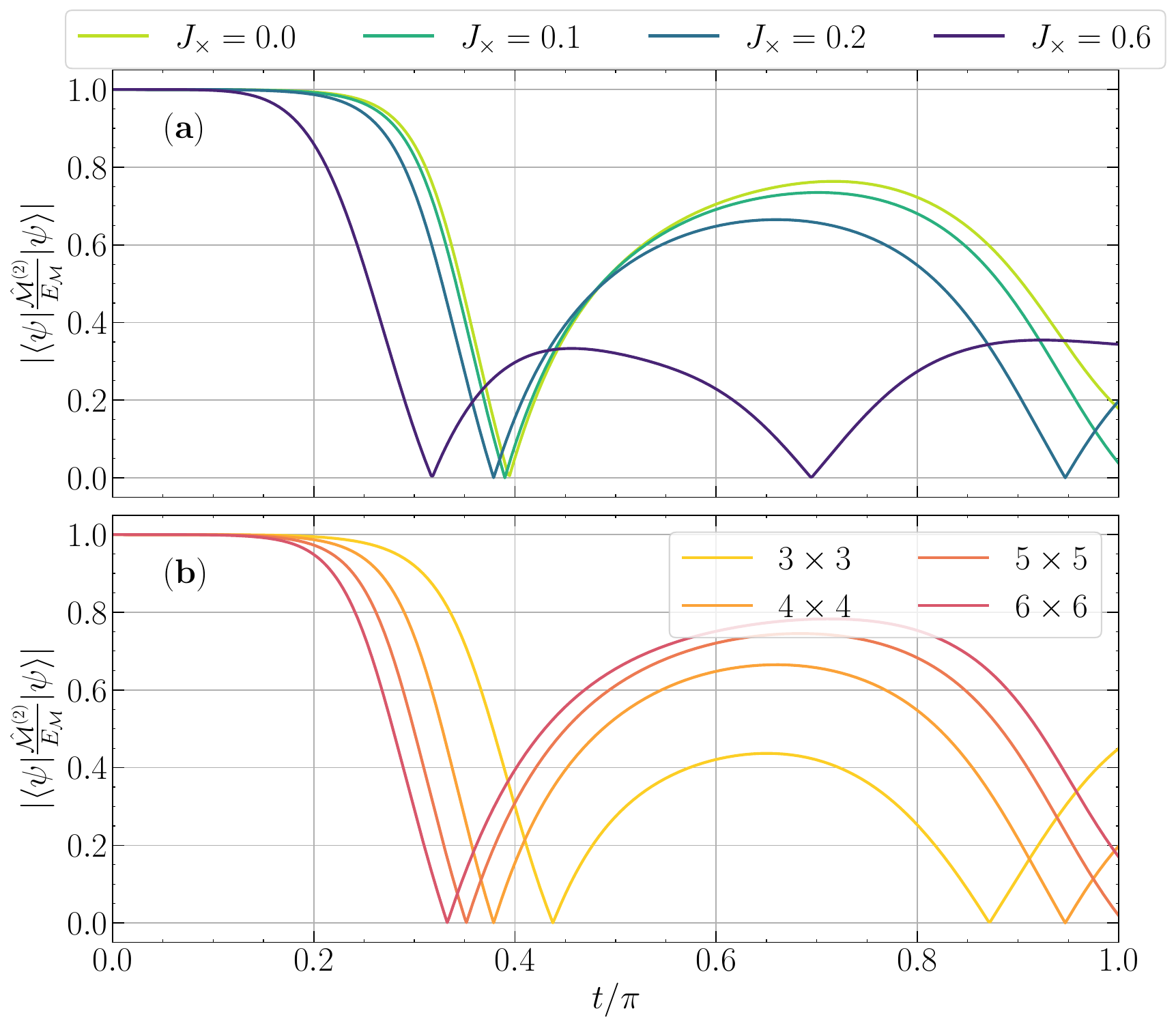}
\caption{
(a) Comparison of the overlap between \(|\psi(t)\rangle\) and the normalized \(\hat{\mathcal{M}}(t)|\psi(t)\rangle\) under the second-order emergent Hamiltonian $\hat{\mathcal{M}}^{(2)}$, for different diagonal hoppings  $J_\times$ in $\hat H_f$, on a $6 \times 6$ lattice. The initial state consists of three particles at the lower corner of the square lattice, specifically at sites \((0,0)\), \((1,0)\), and \((0,1)\). (b) The same as in (a), but contrasting different system sizes for $J_\times = 0.2$. 
}
\label{fig:fig_5}
\end{figure}

To verify their effectiveness, we report in Fig.~\ref{fig:fig_4} the normalized overlap between the time-evolved state $|\psi(t)\rangle$ and $\hat {\mathcal M}^{(1,2)}(t)|\psi(t)\rangle$, for an initial state consisting of two particles positioned at opposite corners of the $L_x\times L_y$ lattice~\footnote{Alternatively, one often employs a different metric for the measure of the appropriateness of $\hat {\mathcal M}^{(n)}(t)$ as an accurate emergent Hamiltonian, such as computing the overlap of the desired eigenstate of the latter with the time-evolved state: $|\langle \Psi_t|\psi(t)\rangle|$~\cite{Vidmar2017, Zhang2021}. This is facilitated by the fact that, as a unitary transformation of $\hat H_0$, $\hat {\mathcal M}(t)$ preserves its spectrum. In our case, owing to the form of $\hat H_0 = \sum_{\bf l}(l_x+l_y)\hat a_{\bf l}^\dagger a_{\bf l}^{\phantom{\dagger}}$, and the chosen initial state $|\psi(0)\rangle$, a symmetric excitation of two excitations at opposite corners of the lattice, there is an extensive degeneracy, complicating the task of monitoring the overlap by inspecting the corresponding eigenvalue to infer what $|\Psi_t\rangle$ is.}, for $J_\times = 0$. That is, we normalize this overlap by $E_{\hat {\mathcal M}}\equiv||\hat {\mathcal M}(t)|\psi(t)\rangle||$: if $\hat {\mathcal M}(t)$  were exact, \(\hat{\mathcal{M}}(t)|\psi(t)\rangle\) would be a scalar multiple of \(|\psi(t)\rangle\), making the overlap $\langle \psi(t)|\hat {\mathcal M}(t)|\psi(t)\rangle/E_{\hat {\mathcal M}}=1$. Since \(\hat{\mathcal{M}}(t)\) is an approximation in our case, the deviation of this overlap from one quantifies the accuracy of this approximation, indicating how efficiently \(\hat{\mathcal{M}}(t)\) preserves \(|\psi(t)\rangle\) at a given time \(t\). Concretely, the normalization would be time independent and equal to the absolute value of the energy associated with the initial state in $\hat H_0$: $E_{\hat {\mathcal M}} =  |\langle\psi(t)|\hat {\mathcal M}|\psi(t)\rangle|=|\langle \psi(t=0)|\hat H_0|\psi(t=0)\rangle|$, where we make use of the definition of the \textit{exact} emergent Hamiltonian.

The results show that the overlap is close to one at short timescales, as expected, and remains so for longer $t$ when considering the higher-order approximation $\hat {\cal M}^{(2)}(t)$ instead of $\hat {\cal M}^{(1)}(t)$---Figs.~\ref{fig:fig_4}(a) and \ref{fig:fig_4}(b), with the same vertical scale. Furthermore, increasing the system size helps maintain the overlap near one for larger timescales, although it eventually decreases steadily. Additionally, a secondary approach can be used, given the form of our initial state: at short times, one effectively has the dynamics of independent particles, since they are far apart in the lattice. As such, a potentially better emergent Hamiltonian is to take the one for a single particle, in which the spin-mapping is possible, and promote it to the many-body setting---that is, using $\hat {\cal M}(t)$ in Eq.~\eqref{eq:M_NN} as a "hopping matrix" and applying it directly to the many-body regime, thereby discarding all nonlocal interactions present in the full emergent Hamiltonian. We refer to this approximation as $\hat{\mathcal{M}}_{\text{spin}}(t)$, owing to the exact spin mapping in the single-particle case.

Remarkably, Fig.~\ref{fig:fig_4}(c) shows that this approximate emergent Hamiltonian (note the different vertical scale) is substantially more effective in preserving the time-dependent state $|\psi(t)\rangle$ in comparison to low-order truncated ones. One of the reasons for that stems from an inherent timescale in the problem for this case: Due to the spin mapping in the single-particle case, we expect a periodicity of $2\pi$, implying that a single excitation should take a time $2\pi$ to traverse the entire lattice and return. Consequently, it takes approximately $\pi$ for a particle to travel from one corner to the opposite corner, and $\pi/2$ for two particles initially at opposite corners to effectively "meet". Thus, for $t \lesssim \pi/2$, the dynamics are in practice single-particle-like, during which $\hat {\cal M}_{\text{spin}}$ remains a good approximation. Beyond this time, interactions (hard-core constraint) are unavoidable, introducing deviations due to the approximate nature of $\hat{\mathcal{M}}(t)$---the inset in Fig.~\ref{fig:fig_4}(a) contrasts the different approximate emergent Hamiltonians at short times, showing how $\hat {\cal M}_{\rm spin}$ is more robust in preserving the overlap. This effect is further enhanced in larger lattices, where the picture of independent particles at short times becomes considerably more reliable.
\begin{figure*}[t] 
  \includegraphics[width=0.9\textwidth]{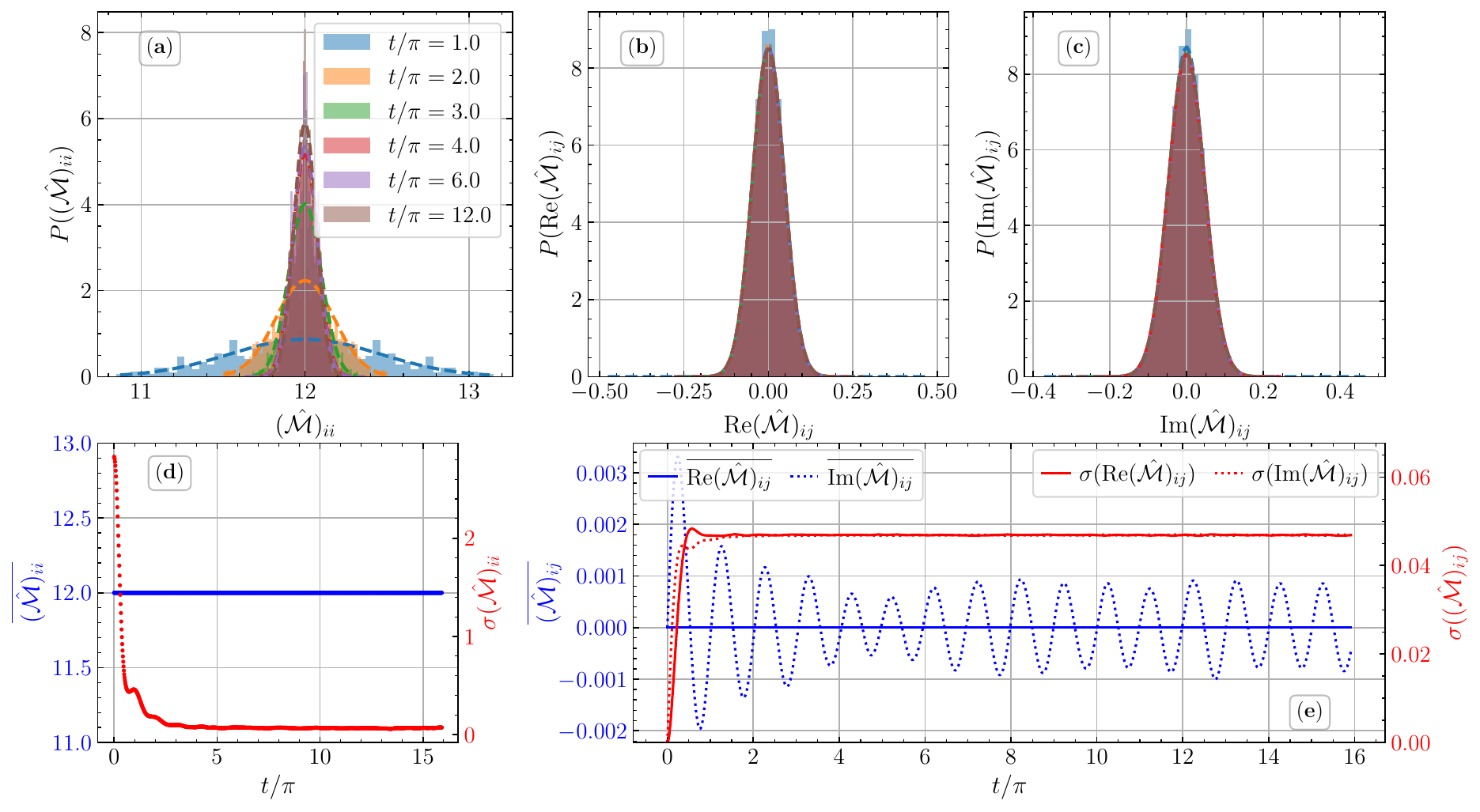}
    \caption{Probability density of occurrence of the emergent Hamiltonian $\hat {\cal M}(t)$ matrix elements in the Fock basis: (a) diagonal elements, and the (b) real and (c) imaginary parts of the off-diagonal elements at different time stamps; Gaussian fittings are shown with dashed lines. (d) and (e) Dynamics of the mean and standard deviation of the fitted Gaussian for the diagonal (d) and off-diagonal (e) matrix elements. All data refer to the exact $\hat {\cal M}(t)$ for the model in Eq.~\eqref{eq:Hf_NNN} with $\hat{H}_0 = \sum_{\bf l}\left(l_x+l_y\right)\hat{a}_{\bf l}^{\dagger}\hat{a}_{\bf l}^{\phantom{\dagger}}$ on a $4\times 4$ lattice with three excitations and a constant diagonal hopping amplitude $J_{\times} = 0.6$.}  
    \label{fig:fig_6}
    \end{figure*}
Going beyond the two-particle case, we now contrast the effect of a homogeneous $J_\times$ in the overlap for the case where one has an initial state with three particles in one corner of the lattice, minimizing $\langle \psi(0)|\hat H_0|\psi(0)\rangle$. We focus on the truncated second-order emergent Hamiltonian $\hat {\cal M}^{(2)}(t)$, since no approximate direct mapping to a spin-like emergent Hamiltonian is possible, and, due to the nature of our initial state, the hard-core constraint sets in right from the start. The larger the number of hopping terms in $\hat H_f$, the more likely it is that a low-order truncation $\hat {\cal M}^{(n)}(t)$ would not preserve the state---this can be understood in terms of additional corrections the $n$-order emergent Hamiltonian would exhibit, leading to an effectively poor approximation.

This is indeed observed in Fig.~\ref{fig:fig_5}(a), where the overlap systematically decays from one at shorter times as $J_\times$ is increased. In fact, since some of these corrections lead to a renormalization of the onsite energies, the larger the lattice, the worse one would expect that $\hat {\cal M}^{(n)}$ would fare in storing the time-dependent state, as is indeed observed in Fig.~\ref{fig:fig_5}(b).

\section{Spectral Analysis} \label{sec:spectral}

As we have noted, the emergent Hamiltonian is, by definition, a unitary transformation of \( \hat{H}_0 \):
\begin{equation}
    \hat{\mathcal{M}}(t) = \hat U \hat{H}_0 \hat U^{\dagger},
    \label{eq:exact_M}
\end{equation}
where \( \hat U \equiv e^{-{\rm i}\hat H_f t} \) at a given time \( t \). This ensures that the eigenvalues of \(\hat{\mathcal{M}}(t)\) are identical to those of \( \hat{H}_0 \). Since \( \hat{H}_0 \) is clearly integrable for the cases we have investigated so far, \(\hat{\mathcal{M}}(t)\) does not exhibit any chaotic behavior as identified by the emergence of level repulsion~\cite{Oganesyan2007, Atas2013, D'Alessio2016}. In fact, for our choice of \( \hat{H}_0 \), the eigenvalues are evenly spaced with integer-valued gaps.

Despite this, the structure of \(\hat{\mathcal{M}}(t)\) becomes increasingly non-local as higher-order terms in its nested commutator expansion are included. These terms include density-assisted hoppings (see Appendix \ref{sec:appA}) for second order, for example, where the hopping amplitude depends on the occupation of additional sites. With each successive order, the range of sites contributing to this effect extends further. This behavior arises directly from the hard-core constraint, distinguishing the system from free bosons or fermions. In general, the commutator of two quadratic terms composed solely of free bosonic or fermionic creation and annihilation operators remains quadratic, preventing the emergence of density-assisted (four-body or higher) terms~\footnote{This can be immediately seen for the case of fermions ($\hat c$ and $\hat c^\dagger$ operators) or bosons ($\hat b$ and $\hat b^\dagger$ operators), where $\bigl[\hat a_i^{\dagger}\hat a_j,\;\hat a_k^{\dagger}\hat a_\ell\bigr]
= \delta_{j k}\,\hat a_i^{\dagger}\hat a_\ell
- \delta_{i \ell}\,\hat a_k^{\dagger}\hat a_j$, where $\hat a_i = \hat c_i$ or $\hat b_i$. For hard-core bosons, on the other hand, this commutator leads to $\bigl[\hat a_i^{\dagger}\hat a_j,\;\hat a_k^{\dagger}\hat a_\ell\bigr]
= \delta_{j k}\,
  \hat a_i^{\dagger} \bigl(1-2\hat n_j\bigr) \hat a_\ell
- \delta_{i \ell}\,
  \hat a_k^{\dagger} \bigl(1-2\hat n_i\bigr) \hat a_j$.}. However, the hard-core constraint introduces strong interactions, leading to the formation of these inherently nonlocal terms.

As additional terms are incorporated, \(\hat{\mathcal{M}}(t)\) becomes increasingly dense, raising the suspicion that it may effectively resemble a random matrix. In the infinite-order limit (for any finite \(t\)), \(\hat{\mathcal{M}}(t)\) is highly nonsparse and appears ``random.'' This appearance is, however, misleading: \(\hat{\mathcal{M}}(0)=\hat{H}_0\) is diagonal in the Fock basis, and the growth of nonlocal terms with time merely creates the semblance of randomness. Because \(\hat{\mathcal{M}}(t)\) is unitarily similar to \(\hat{H}_0\) (and thus Hermitian with the same eigenvalues), the spectrum remains highly regular and does not exhibit Wigner-Dyson level repulsion. This dichotomy between random-matrix-like structure without level repulsion is explored in Fig.~\ref{fig:fig_6}, where we examine histograms of the matrix elements of the numerically exact emergent Hamiltonian $\hat{\mathcal{M}}(t)$ [Eq.~\eqref{eq:exact_M}] at different time stamps. This is constructed for $\hat H_f$ in Eq.~\eqref{eq:Hf_NNN}, and $\hat{H}_0 = \sum_{\bf l}\left(l_x+l_y\right)\hat{a}_{\bf l}^{\dagger}\hat{a}_{\bf l}^{\phantom{\dagger}}$ for the three excitation subspace.

We start with the (normalized) histograms of diagonal elements [Fig.~\ref{fig:fig_6}(a)], which are real, where they rapidly develop a Gaussian structure, overcoming the structured pattern associated with $\hat H_0$ at short times. Notably, the mean is time independent, a direct consequence of the trace conservation under unitary transformations. Meanwhile, such a Gaussian structure exhibits a standard deviation that equilibrates to a saturation value [Fig.~\ref{fig:fig_6}(d)]. In turn, for off-diagonal elements [the real and imaginary parts, Fig.~\ref{fig:fig_6}(b) and \ref{fig:fig_6}(c), respectively], the pattern is similar, but their mean hovers around zero, and the standard deviation quickly grows from zero to an equilibrium value [Fig.~\ref{fig:fig_6}(e)]. 

To characterize the similarity of $\hat {\cal M}(t)$ to a random matrix at large $t$, we quantify the fluctuations of matrix elements by the ratio \( r \equiv \sigma_{\mathrm{off}}/\sigma_{\mathrm{diag}} \), where \(\sigma_{\mathrm{off}}\) denotes the standard deviation of the off-diagonal elements and \(\sigma_{\mathrm{diag}}\) that of the diagonal elements. Numerically, the real and imaginary parts of the off-diagonal elements coincide \((\sigma_{\mathrm{off}}^{\mathrm{real}}\simeq \sigma_{\mathrm{off}}^{\mathrm{imag}})\), thus we treat them jointly. For the \(4\times4\) example in Fig.~\ref{fig:fig_6}, we find a representative late-time value \( r \approx 0.63 \), consistent across the sampled time stamps. Across system sizes and fillings $f$, and excluding two dilute-filling outliers (\(5\times5, f=3/25 = 0.125\) and \(5\times5, f =0.08\)), the ratio has mean $\overline r=0.662$ with sample standard deviation \(0.029\), spanning the range \(0.616\)-\(0.697\). At first sight, they look suspiciously close to random-matrix expectations. For comparison, a Gaussian ensemble within random matrix theory (RMT), with standard normalization, gives $r_{\mathrm{RMT}}=1/\sqrt{2}\approx0.707$~\cite{D'Alessio2016}. Our slightly lower values are plausibly explained by the finite Hilbert spaces accessible here and by residual symmetries such as particle-hole symmetry that constrain the matrix elements. Even though the emergent Hamiltonian is dense and includes most nonlocal terms, it is still a unitary transformation of \(\hat H_0\) rather than a draw from an independent Gaussian ensemble.

\section{Another example: Storing a GHZ state}
\label{sec:oat_dicke}

We now move on to yet another example of using the emergent Hamiltonian 
protocol aimed at preparing and storing a GHZ state, which exhibits fragile, globally distributed entanglement. Such correlations are effectively impossible to maintain in MBL-like protocols, which tend to drive the system toward a classical mixture of its macroscopic branches due to their inherent locality. In the context of quantum circuits, Ref.~\cite{Song2019} shows that, in place of the conventional $O(N)$-depth \textsc{cnot} ladder used for GHZ generation, a constant-depth $O(1)$ can be realized through a single collective entangling operation combined with global rotations (i.e., all-to-all couplings among qubits) and parallel parity readout for verification. When coupled with our emergent Hamiltonian protocol, which freezes the complete state including its global correlations, this approach provides a direct route to preparing and storing GHZ states.

Following Refs.~\cite{Song2019, Xu2020, Xu2022}, we focus on the collective "one-axis twisting" (OAT) setting, also emulated with trapped atoms~\cite{Gross2010}, where the entangling Hamiltonian is~\cite{Kitagawa1993}
\begin{equation}
\hat H_{\rm OAT}=-\lambda\hat S_z^2.
\label{eq:H_OAT_dicke}
\end{equation}
We restrict the dynamics to the totally symmetric spin sector of $L$ qubits, $s=L/2$, where the Dicke basis states $\{|s\ m\rangle\}$ are defined as eigenstates of the global spin operators
\begin{equation}
\hat S^2|s\ m\rangle = s(s+1)|s\ m\rangle\quad \text{and}\quad \hat S_z|s\ m\rangle=m|s\ m\rangle\ ,
\end{equation}
with $m=-s,-s+1,\dots,s$. We start from the spin-coherent state aligned along the $-y$ axis in the Bloch sphere,
\begin{equation}
|\psi(0)\rangle = \Bigl(\frac{|0\rangle-{\rm i}|1\rangle}{\sqrt{2}}\Bigr)^{\otimes L},
\end{equation}
or, equivalently, in the Dicke basis,
\begin{equation}
\ket{\psi(0)}=\sum_{m=-s}^sc_m|m\rangle,\quad
c_m=2^{-L/2}(-{\rm i})^{m+s}\sqrt{\binom{L}{m+s}},
\end{equation}
which is an eigenstate of $\hat H_0=\hat S_y$. What we aim to show is that an emergent Hamiltonian $\hat {\cal M}(t) = e^{-{\rm i}\hat H_{\rm OAT}t}\hat S_y e^{+{\rm i}\hat H_{\rm OAT}t} = e^{+{\rm i}\lambda \hat S_z^2t}\,\hat S_y\,e^{-{\rm i}\lambda \hat S_z^2 t}$ has a closed form that can be used to freeze desired states. In particular, within the subspace defined by $\{|m\rangle\}$, this can be simplified by decomposing $\hat S_y$ into the spin ladder operators, $\hat S_y = \frac{\hat S_+ - \hat S_-}{2{\rm i}}$, and noticing that $e^{+{\rm i}\lambda \hat S_z^2t}\,\hat S_\pm\,e^{-{\rm i}\lambda \hat S_z^2 t}|m\rangle = e^{{\rm i}\lambda t[(m\pm 1)^2-m^2]}\hat S_\pm|m\rangle = e^{{\rm i}\lambda t(\pm 2m+1)}\hat S_{\pm}|m\rangle$. As a result, the emergent Hamiltonian assumes a compact form,
\begin{equation}
    \hat {\cal M}(t) = \frac{1}{2{\rm i}}\left[e^{{\rm i}\lambda t (2\hat S_z+1)}\hat S_+ - e^{-{\rm i}\lambda t(2\hat S_z-1)}\hat S_-\right]\ .
\end{equation}
Alternatively, we can also use the finite matrix elements of the operators in the $\{|m\rangle\}$ basis, $(S_z)_{mm} = m$ and $(S_\pm)_{m,m\pm1} = \sqrt{s(s+1)-m(m\pm1)}$, to write the matrix elements of the emergent Hamiltonian as

\begin{equation}
    \left(\hat{\mathcal M}(t)\right)_{m,m\pm 1}
=
\mp \frac{1}{2i}\ e^{\mp \rm i\lambda t (2m \pm 1)}\sqrt{s(s+1)-m(m\pm1)}
    \label{eq:M_OAT}
\end{equation}

This makes it explicit that $\hat{\cal M}(t)$ is tridiagonal in the Dicke basis, and to note that the $m$-dependent phases $e^{\pm {\rm i}\lambda t(2m\pm 1)}$ appearing in $\hat{\cal M}(t)$ become particularly simple at special times. 

For instance, at $t=\pi/(2\lambda)$, one finds $e^{{\rm i}(\pi/2)(2m+1)} = {\rm i}(-1)^m$, producing an alternating sign pattern between neighboring Dicke sectors. This ``staggered hopping'' frustrates intermediate $m$-values, leading to an eigenstate that approaches a GHZ state, for large $L$, along the $y$-axis on the equator,
\begin{align}
|{\rm GHZ}^y_\phi\rangle&=\frac{\ket{0_y\ldots0_y}+e^{{\rm i}\phi}\ket{1_y\ldots1_y}}{\sqrt{2}}\notag\\
&=\hat R_x(-\pi/2)\ket{{\rm GHZ}_\phi},
\end{align}
where $\ket{\mathrm{GHZ}_\phi}$ is the symmetric superposition of the $m=\pm s$ states, and $R_x(\theta)$ a global rotation about the $x$-axis by $\theta$. At $t=\pi/\lambda$, on the other hand, all phases align ($e^{{\rm i}\pi(2m+1)}=-1$), recovering the original $\hat S_y$ up to a global sign and reviving the initial spin-coherent state. These special times thus correspond to the well-known ``cat-state'' and ``revival'' points of the one-axis twisting dynamics~\cite{Gross2010, Song2019} and demonstrate why $\hat{\mathcal M}(t)$ can be used to freeze a macroscopically entangled GHZ-like state.

\begin{figure}[t!]
  \includegraphics[width=\columnwidth]{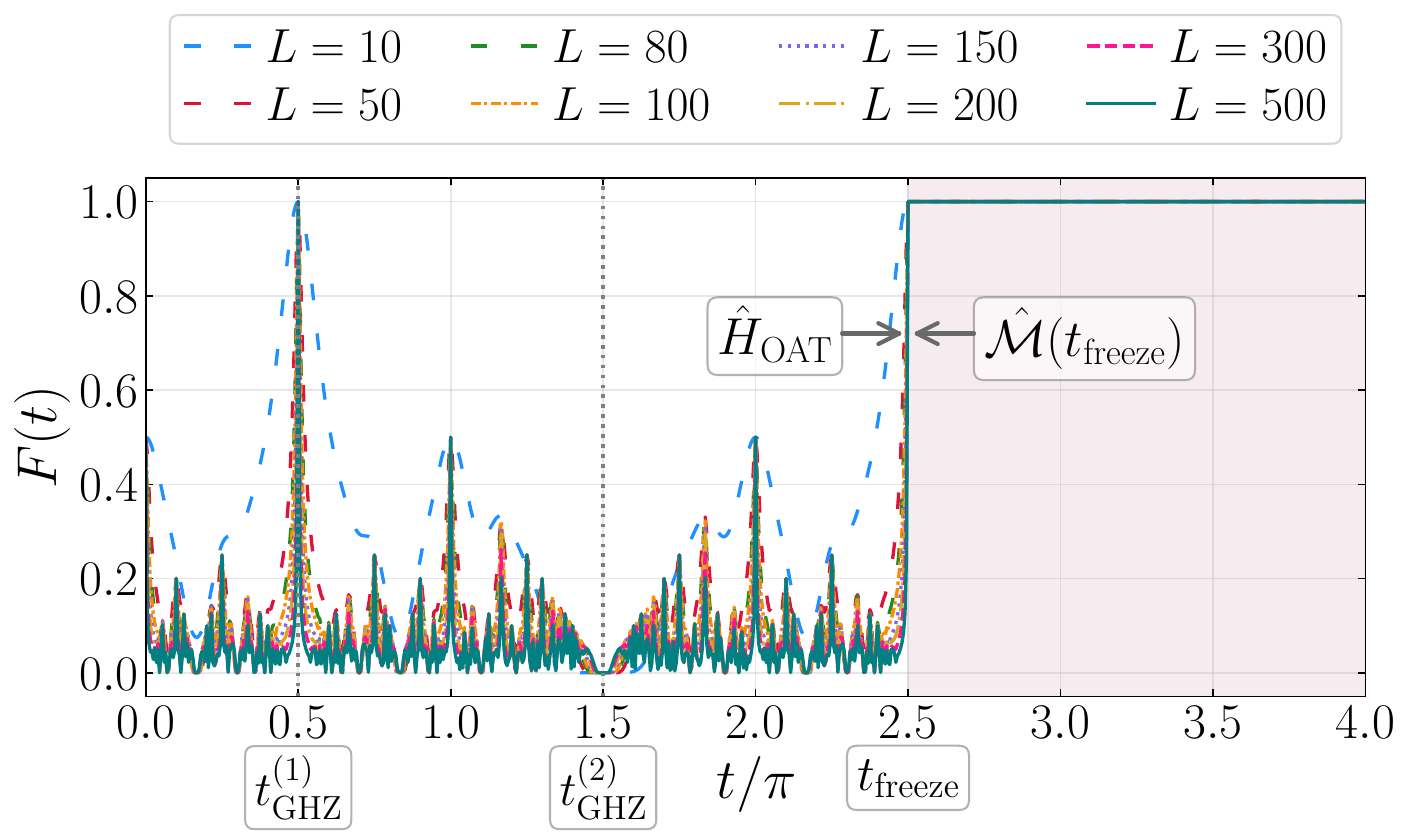}
  \caption{Fidelity $F(t)=\big|\langle \mathrm{GHZ}_{3\pi/2}|\,R_x(-\frac{\pi}{2})\,\psi(t)\rangle\big|^2$ between the rotated state $R_x(-\pi/2)\ket{\psi(t)}$ and the target GHZ at $t^{(1)}_{\rm GHZ}=\pi/(2\lambda)$ for various $L$. The dotted vertical lines indicate $t^{(1,2)}_{\rm GHZ}$.}
  \label{fig:fidelity_ghz_freeze}
\end{figure}

Let us now show how the protocol works in practice. By setting $\lambda=1$, at times $t^{(1)}_{\rm GHZ}=\pi/2$ (modulo $2\pi$), the evolution produces the GHZ state  
\begin{equation}
|\psi(t=\pi/2)\rangle=|\text{GHZ}_{\phi=3\pi/2}\rangle=\frac{1}{\sqrt{2}}\left(|0\cdots0\rangle - {\rm i}\,|1\cdots1\rangle\right)\ ,
\end{equation}
while at times $t^{(2)}_{\rm GHZ}=3\pi/2$ (modulo $2\pi$) it yields another GHZ state
\begin{equation}
|\psi(t=3\pi/2)\rangle = |\text{GHZ}_{\phi=\pi/2}\rangle=\frac{1}{\sqrt{2}}\left(|0\cdots0\rangle + {\rm i}\,|1\cdots1\rangle\right)\ .
\end{equation}
These two GHZ states differ by the relative phase $\pm {\rm i}$ between the $|0\cdots0\rangle$ and $|1\cdots1\rangle$ components and are mutually orthogonal.

Because the GHZ produced via this protocol naturally lies along the $y$ direction of the collective Bloch sphere, we rotate the time-evolved state $|\psi(t)\rangle$ about the $x$ axis by $-\pi/2$ to map the equatorial GHZ to the computational GHZ state before calculating the fidelity:
\begin{equation}
F(t)=\left|\langle {\rm GHZ}_{\phi}|{\hat R_x(-\pi/2)\psi(t)\rangle}\right|^2 \ .
\end{equation}

Figure~\ref{fig:fidelity_ghz_freeze} shows the fidelity dynamics generated by $\hat H_{\rm OAT}$ starting from the $-y$-oriented spin coherent state, $|1_y\rangle^{\otimes L}$ for different numbers of qubits $L$. The curve peaks at $t^{(1)}_{\rm GHZ}$ modulo $2\pi$, where the GHZ forms. To attest the efficiency of our quantum memory protocol, we quench the dynamics at time $t_{\rm freeze}=t^{(1)}_{\rm GHZ}+2\pi$ by switching its generator from $\hat H_{\rm OAT}$ to the emergent Hamiltonian $\hat{\cal M}(t_{\rm freeze})$, thus freezing this globally distributed entangled state indefinitely.

\section{Summary and Outlook} \label{sec:conclusion}
The quest to store highly entangled quantum states, thereby realizing a robust quantum memory, is one of the fundamental cornerstones of the quantum computing paradigm. In this work, we have demonstrated that the protocol of emergent Hamiltonians can achieve this goal in several contexts, in particular for stabilizing entangled states such as tensor products of Bell states or GHZ states. The essential ingredient is the focus on compact emergent Hamiltonians, which can be readily implemented within existing quantum computing platforms.

In situations where no compact form exists, such as in the many-body regime in more than one dimension, approximate emergent Hamiltonians can be derived to freeze the corresponding entangled state after an initial short dynamics effectively. From a mathematical perspective, we further revealed that such Hamiltonians inherit quantum chaotic features: while they arise from simple parent Hamiltonians with non-entangled eigenstates, their unitary transformation often generates dense operator structures reminiscent of chaotic systems. In this sense, their matrix properties align with the corresponding random matrix class, even though their spectra do not display level repulsion, distinguishing them from genuinely chaotic ensembles.

An intriguing direction for future exploration concerns ergodicity. The eigenstate thermalization hypothesis (ETH) asserts that the expectation values of few-body observables in energy eigenstates are smooth functions of the eigenenergy, with eigenstate-to-eigenstate fluctuations that vanish exponentially with system size~\cite{Rigol2008, D'Alessio2016, Mondaini2016}. Whether generic emergent Hamiltonians satisfy ETH remains an open question. A key subtlety is that a local observable in the original basis may be mapped to a highly nonlocal one under the unitary transformation defining $\hat{\cal M}(t)$, thereby evading the conventional ETH framework. Systematic tests across a large class of emergent Hamiltonians could clarify the extent to which ergodicity and ETH hold (or fail) in this setting.

Finally, concerning experimental explorations, one limiting characteristic of actual devices is the inherent presence of decoherence, which can ultimately spoil the ideal quantum memory enabled by the emergent Hamiltonian. Incoherent processes such as relaxation and dephasing progressively suppress phase coherence and redistribute population, thereby degrading eigenstates of $\hat{\cal M}(t)$. A quantitative analysis of these effects for the GHZ case is presented in Appendix~\ref{sec:noise}.

Nonetheless, \emph{coherent} scrambling arising from imperfect control or residual interactions is expected to occur on a time scale that scales inversely with the energy gap. Since these gaps can be tuned through appropriate manipulation of the parent Hamiltonian $\hat H_0$, one can prolong the operational memory window: large spectral gaps intrinsically suppress coherent (unitary) perturbations by energetically isolating the target eigenstate.


It is worth emphasizing that the emergent-Hamiltonian framework naturally acts as a design toolbox: the parent Hamiltonian \(\hat H_{0}\) fixes the eigenspectrum, while the driving Hamiltonian \(\hat H_{f}\) shapes the dynamics. This two-Hamiltonian structure enables one to engineer both preparation and storage of entangled states, ensuring that a desired state can be made an eigenstate of \(\hat H_{0}\) and stabilized against coherent perturbations. In this way, the framework provides a flexible and experimentally relevant route toward the controlled realization of long-lived quantum memories.
\vskip0.1in

\begin{acknowledgments}R.M.~acknowledges support from the T$_c$SUH Welch Professorship Award. Numerical simulations were performed with resources provided by the Research Computing Data Core at the University of Houston. This work also used TAMU ACES at Texas A\&M HPRC through allocation PHY240046 from the Advanced Cyberinfrastructure Coordination Ecosystem: Services \& Support (ACCESS) program, which is supported by U.S. National Science Foundation grants 2138259, 2138286, 2138307, 2137603, and 2138296.
\end{acknowledgments}

\section*{Data Availability}

The data that support the findings of this article are openly available~\cite{zenodo}, embargo periods may apply.

\bibliography{references}

\appendix
\onecolumngrid
\section{Approximate Emergent Hamiltonians in the many-body setting for two-dimensional systems}
\label{sec:appA}
In the main text, Sec.~\ref{sec:many_particle}, we argued that tackling the many-body case in more than one dimension leads to an elusive compact form for the emergent Hamiltonian, owing to the impossibility of mapping the problem to large spins. As such, one needs to rely on a truncation of Eq.~\eqref{eq:emergent_Hamiltonian}, as done in other Hamiltonians~\cite{Vidmar2017, Zhang2021}, whose truncation order $n$ introduces an error ${\cal O}(t^n)$. In what follows, we explicitly compute the functional form of these approximate emergent Hamiltonians in two-dimensional lattices, focusing both on the case where one has only nearest-neighbor (NN) hoppings as well as when a homogeneous next-nearest-neighbor (NNN) hopping is included.

\subsection{Two-dimensional nearest-neighbor approximate emergent Hamiltonian---Many excitations}
Our starting point is the entangling Hamiltonian $\hat H_f$ and the initial Hamiltonian $\hat H_0$, given by
\begin{align}
\hat{H}_f
&= \sum_{\mathbf l}\frac{\sqrt{l_x\!\left(L_x-l_x\right)}}{2}
   \Bigl(\hat a_{\mathbf l+\hat{x}}^{\dagger}\hat a_{\mathbf l}^{\phantom{\dagger}}+\mathrm{H.c.}\Bigr)
 + \sum_{\mathbf l}\frac{\sqrt{l_y\!\left(L_y-l_y\right)}}{2}
   \Bigl(\hat a_{\mathbf l+\hat{y}}^{\dagger}\hat a_{\mathbf l}^{\phantom{\dagger}}+\mathrm{H.c.}\Bigr), \text{ and} 
\\[4pt]
\hat{H}_0
&= \sum_{\mathbf l}\!\left(l_x+l_y\right)
   \hat a_{\mathbf l}^{\dagger}\hat a_{\mathbf l}^{\phantom{\dagger}}\ .
\end{align}
Computing the truncated emergent Hamiltonian up to first order, which involves computing the commutator between $\hat H_0$ and $\hat H_f$, reads
\begin{align}
\hat{\mathcal M}^{(1)}_{\rm NN}(t)
&= \hat H_0 - \mathrm{i}\,t\,\hat{\mathcal H}_1 \notag\\
&= \sum_{\mathbf l}\!\left(l_x+l_y\right)\hat a_{\mathbf l}^{\dagger}\hat a_{\mathbf l}^{\phantom{\dagger}}
 - \frac{t}{2}\Biggl[
   \sum_{\mathbf l} \sqrt{l_x\!\left(L_x-l_x\right)}
   \Bigl\{\mathrm{i}\,\hat a_{\mathbf l}^{\dagger}\hat a_{\mathbf l+\hat{x}}^{\phantom{\dagger}}+\mathrm{H.c.}\Bigr\}
 + \sum_{\mathbf l} \sqrt{l_y\!\left(L_y-l_y\right)}
   \Bigl\{\mathrm{i}\,\hat a_{\mathbf l}^{\dagger}\hat a_{\mathbf l+\hat{y}}^{\phantom{\dagger}}+\mathrm{H.c.}\Bigr\}
 \Biggr]\ .
 \label{eq:app_M_1_NN}
\end{align}
In the calculation of $\hat {\cal H}_1$, we make use of the hard-core boson commutation identities $[\hat a_{\bf i},\hat n_{\bf j}] =\delta_{\bf ij}\hat a_{\bf i}$ and $[\hat a_{\bf i}^\dagger,\hat n_{\bf j}] =-\delta_{\bf ij}\hat a_{\bf i}^\dagger$, where $\hat n_{\bf i} = \hat a_{\bf i}^\dagger\hat a_{\bf i}^{\phantom{\dagger}}$ is the number operator, which help evaluate commutators of the form $[\hat a_{\bf i}^\dagger \hat a_{\bf j}^{\phantom{\dagger}}, \sum_{\bf l}w_{\bf l}\hat n_{\bf l}] = (w_{\bf j}-w_{\bf i})\hat a_{\bf i}^\dagger\hat a_{\bf j}^{\phantom{\dagger}}$. Given the nearest neighbor structure for the hopping terms in $\hat H_f$, this means that $w_{\bf j}-w_{\bf i} = w_{{\bf l}+\hat x} - w_{\bf l} = w_{{\bf l}+\hat y} - w_{\bf l} = 1$ for the $\hat H_0$ highlighted above. Finally, the physical interpretation of this first-order correction is to generate terms of current-like form, whose amplitude is modulated by the time $t$. 

Proceeding to the second-order truncation, we compute
\begin{align}
\hat{\mathcal M}^{(2)}_{\rm NN}(t)
&= \hat H_0 - \mathrm{i}\,t\,\hat{\mathcal H}_1 - \frac{t^2}{2}\,\hat{\mathcal H}_2 \notag\\
&= \sum_{\mathbf l}\!\left(l_x+l_y\right)\hat a_{\mathbf l}^{\dagger}\hat a_{\mathbf l}^{\phantom{\dagger}}
 - \frac{t}{2}\Biggl[
 \sum_{\mathbf l} \sqrt{l_x\!\left(L_x-l_x\right)}
   \Bigl\{\mathrm{i}\,\hat a_{\mathbf l}^{\dagger}\hat a_{\mathbf l+\hat{x}}^{\phantom{\dagger}}+\mathrm{H.c.}\Bigr\}
 + \sum_{\mathbf l} \sqrt{l_y\!\left(L_y-l_y\right)}
   \Bigl\{\mathrm{i}\,\hat a_{\mathbf l}^{\dagger}\hat a_{\mathbf l+\hat{y}}^{\phantom{\dagger}}+\mathrm{H.c.}\Bigr\}
 \Biggr] \notag\\[2pt]
&\quad
 - \frac{t^2}{4}\Biggl(
 \sum_{\mathbf l}\!(l_x\!-\!1)(L_x\!-\!l_x\!+\!1)\,\hat a_{\mathbf l}^{\dagger}\hat a_{\mathbf l}^{\phantom{\dagger}}
 - \sum_{\mathbf l}\!l_x(L_x\!-\!l_x)\,\hat a_{\mathbf l}^{\dagger}\hat a_{\mathbf l}^{\phantom{\dagger}}
 + \sum_{\mathbf l}\!(2l_x\!-\!L_x\!-\!1)\,\hat a_{\mathbf l}^{\dagger}\hat a_{\mathbf l}^{\phantom{\dagger}}
 \notag\\[-2pt]
&\quad\quad\quad
 + \sum_{\mathbf l}\!(l_y\!-\!1)(L_y\!-\!l_y\!+\!1)\,\hat a_{\mathbf l}^{\dagger}\hat a_{\mathbf l}^{\phantom{\dagger}}
 - \sum_{\mathbf l}\!l_y(L_y\!-\!l_y)\,\hat a_{\mathbf l}^{\dagger}\hat a_{\mathbf l}^{\phantom{\dagger}}
 + \sum_{\mathbf l}\!(2l_y\!-\!L_y\!-\!1)\,\hat a_{\mathbf l}^{\dagger}\hat a_{\mathbf l}^{\phantom{\dagger}}
 \notag\\[-2pt]
&\quad\quad\quad
 + 2\sum_{\mathbf l}\!\!\sqrt{l_x\!\left(L_x-l_x\right)}\sqrt{l_y\!\left(L_y-l_y\right)}
   \Bigl\{
     \hat a_{\mathbf l+\hat{x}+\hat{y}}^{\dagger}\hat a_{\mathbf l+\hat{x}+\hat{y}}^{\phantom{\dagger}}
     \hat a_{\mathbf l+\hat{x}}^{\dagger}\hat a_{\mathbf l+\hat{y}}^{\phantom{\dagger}}
     - \hat a_{\mathbf l}^{\dagger}\hat a_{\mathbf l}^{\phantom{\dagger}}
       \hat a_{\mathbf l+\hat{x}}^{\dagger}\hat a_{\mathbf l+\hat{y}}^{\phantom{\dagger}}
   \Bigr\}
 \notag\\[-2pt]
&\quad\quad\quad
 + 2\sum_{\mathbf l}\!\!\sqrt{l_x\!\left(L_x-l_x\right)}\sqrt{l_y\!\left(L_y-l_y\right)}
   \Bigl\{
     \hat a_{\mathbf l+\hat{x}+\hat{y}}^{\dagger}\hat a_{\mathbf l+\hat{x}+\hat{y}}^{\phantom{\dagger}}
     \hat a_{\mathbf l+\hat{y}}^{\dagger}\hat a_{\mathbf l+\hat{x}}^{\phantom{\dagger}}
     - \hat a_{\mathbf l}^{\dagger}\hat a_{\mathbf l}^{\phantom{\dagger}}
       \hat a_{\mathbf l+\hat{y}}^{\dagger}\hat a_{\mathbf l+\hat{x}}^{\phantom{\dagger}}
   \Bigr\}
 \Biggr) \ ,
\end{align}
where the second-order correction, proportional to $t^2$, results in two types of terms: (1) renormalizations of the on-site energies already present in $\hat H_0$, and (2), four-operator terms, with some compactly written as $\hat n_{\bf l^\prime}\hat a_{{\bf l}+\hat x}^\dagger \hat a_{{\bf l}+\hat y}^{\phantom{\dagger}}$, where ${\bf l^\prime} = \hat {\bf l}$ or ${\bf l}+\hat x + \hat y$, that is, a density-assisted hopping term. In these calculations, we repeatedly employ the fundamental commutator identities for hard-core bosons:
\begin{align}
\bigl[\hat a_{\mathbf i}^\dagger \hat a_{\mathbf j}^{\phantom{\dagger}},
      \hat a_{\mathbf k}^\dagger \hat a_{\mathbf l}^{\phantom{\dagger}}\bigr]
&= \delta_{\mathbf j \mathbf k}\,
   \hat a_{\mathbf i}^\dagger (1-2\hat n_{\mathbf j})
   \hat a_{\mathbf l}^{\phantom{\dagger}}
 - \delta_{\mathbf i \mathbf l}\,
   \hat a_{\mathbf k}^\dagger (1-2\hat n_{\mathbf i})
   \hat a_{\mathbf j}^{\phantom{\dagger}}, \\[3pt]
\bigl[\hat a_{\mathbf i}^\dagger \hat a_{\mathbf j}^{\phantom{\dagger}},
      \hat n_{\mathbf k}\bigr]
&= (\delta_{\mathbf j \mathbf k}-\delta_{\mathbf i \mathbf k})\,
   \hat a_{\mathbf i}^\dagger \hat a_{\mathbf j}^{\phantom{\dagger}}.
\end{align}
These are the building blocks for simplifying the nested commutators
appearing in $\hat{\mathcal H}_2$ into the diagonal and
density-assisted hopping contributions written above.

\subsection{Two-dimensional next-nearest neighbor---Many excitations}
Similar calculations can be carried out in the case where one considers an extra next-nearest neighbor hopping term, here treated as homogeneous as in Eq.~\eqref{eq:Hf_NNN}. The entangling and initial Hamiltonians read:
\begin{align}
\hat H_f
&= \sum_{\mathbf l}\frac{\sqrt{l_x\!\left(L_x-l_x\right)}}{2}
   \Bigl(\hat a_{\mathbf l+\hat{x}}^{\dagger}\hat a_{\mathbf l}^{\phantom{\dagger}}+\mathrm{H.c.}\Bigr)
 + \sum_{\mathbf l}\frac{\sqrt{l_y\!\left(L_y-l_y\right)}}{2}
   \Bigl(\hat a_{\mathbf l+\hat{y}}^{\dagger}\hat a_{\mathbf l}^{\phantom{\dagger}}+\mathrm{H.c.}\Bigr)
 \notag\\
&\quad
 + J_\times\sum_{\mathbf l}\Bigl[
    \bigl(\hat a_{\mathbf l+\hat{y}-\hat{x}}^{\dagger}\hat a_{\mathbf l}^{\phantom{\dagger}}+\mathrm{H.c.}\bigr)
  + \bigl(\hat a_{\mathbf l+\hat{y}+\hat{x}}^{\dagger}\hat a_{\mathbf l}^{\phantom{\dagger}}+\mathrm{H.c.}\bigr)
 \Bigr],
\\[4pt]
\hat H_0
&= \sum_{\mathbf l}\!\left(l_x+l_y\right)\hat a_{\mathbf l}^{\dagger}\hat a_{\mathbf l}^{\phantom{\dagger}}\ .
\end{align}
In this case, up to the first order in the truncation, the emergent Hamiltonian results in
\begin{align}
\hat{\mathcal M}^{(1)}_{\rm NNN}(t)
&= \hat H_0 - \mathrm{i}\,t\,\hat{\mathcal H}_1 \notag\\
&= \hat{\mathcal M}^{(1)}_{\mathrm{NN}}(t)
 + \Bigl\{
   (-\mathrm{i}t)J_\times\bigl[
     \sum_{\bf l}\hat a_{{\bf l}-\hat x+\hat y}^{\dagger}\hat a_{\bf l}^{\phantom{\dagger}}
     - \sum_{\bf l}\hat a_{{\bf l}+\hat x+\hat y}^{\dagger}\hat a_{\bf l}^{\phantom{\dagger}}
   \bigr] + \mathrm{H.c.}
 \Bigr\}\ ,
\end{align}
where $\hat{\mathcal{M}}^{(1)}_{\rm NN}(t)$ is defined in Eq.~\eqref{eq:app_M_1_NN}. This shows that the extra hopping terms lead to current-like terms among the next-nearest neighbor sites.

\section{THE EFFECTS OF ENVIRONMENTAL NOISE}
\label{sec:noise}

The Emergent Hamiltonian protocol described in Secs.~\ref{sec:1d}--\ref{sec:oat_dicke} is exact at the level of unitary dynamics: if the system is evolved under an entangling Hamiltonian $\hat H_f$ up to a time $t_{\rm freeze}$ and then quenched to the corresponding emergent Hamiltonian $\hat{\mathcal M}(t_{\rm freeze})$ [Eq.~\eqref{eq:emergent_Hamiltonian} and its closed forms such as Eqs.~\eqref{eq:emergent_1d}, \eqref{eq:M_NN} and \eqref{eq:M_OAT}], the many-body state $|\psi(t_{\rm freeze})\rangle$ is stationary (up to a global phase) because it is an eigenstate of $\hat{\mathcal M}(t_{\rm freeze})$. Actual devices, however, are never perfectly isolated, and the stored state will ultimately degrade due to coupling to uncontrolled environmental degrees of freedom. In this section we quantify how such noise limits the lifetime of the memory, in particular for the GHZ state discussed in Sec.~\ref{sec:oat_dicke}.

We include Markovian energy relaxation and dephasing during the post-quench memory (storage) stage. The preparation dynamics under $\hat H_{\rm OAT}$ is taken to be unitary. Noise is incorporated in the storage stage through coherence times $T_1$ and $T_2$: relaxation is implemented with rate $1/T_1$, while dephasing is included so as to reproduce the prescribed transverse coherence time $T_2$, consistent with the standard decoherence model for superconducting qubits~\cite{Krantz2019}.

All simulations are performed in the natural units of the OAT Hamiltonian $H=-\lambda S_z^2$ with $\hbar=1$, using dimensionless time $t=\lambda_{\rm phys} t_{\rm phys}$. To connect with superconducting-circuit implementations, we consider an interaction strength \( \lambda_{\rm phys}/(2\pi) \approx 2~\mathrm{MHz} \), corresponding to \( \lambda_{\rm phys} = 4\pi \times 10^6~\mathrm{s^{-1}} \), and adopt representative coherence times \( T_1^{\rm phys} = 34~\mu\mathrm{s} \) and \( T_2^{\rm phys} = 20~\mu\mathrm{s} \)~\cite{Song2019}. Expressed in natural units via
\[
T_\alpha = \lambda_{\rm phys} T^{\rm phys}_\alpha,
\]
this yields the dimensionless decay times \( T_1 \approx 136\pi \approx 4.3\times10^2 \) and
\( T_2 \approx 80\pi \approx 2.5\times10^2 \), which are used throughout the analysis. The noisy dynamics are simulated using the permutationally invariant quantum solver (PIQS), which exploits collective symmetry to efficiently evolve the system within the Dicke manifold~\cite{Shammah2018}.

\begin{figure*}[t!]
  \includegraphics[width=0.8\textwidth]{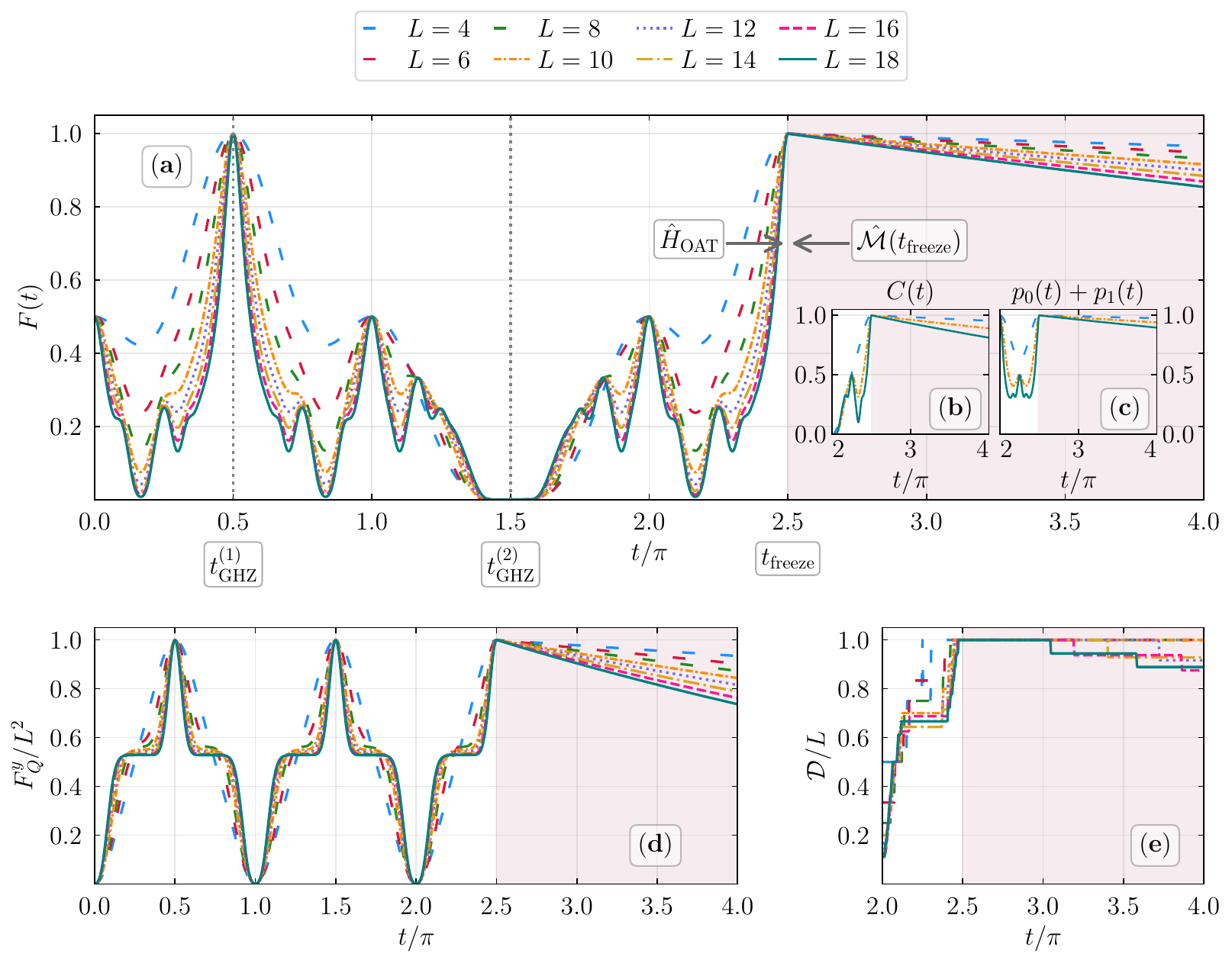}
  \caption{Impact of Markovian relaxation and dephasing on the GHZ-memory protocol of Sec.~\ref{sec:oat_dicke}; we use values of $T_1$ and $T_2$ typical to experiments~\cite{Song2019}. A GHZ state is generated under $\hat H_{\rm OAT}=-\lambda \hat S_z^2$ and then frozen by quenching to $\hat{\mathcal M}(t_{\rm freeze})$ at $t_{\rm freeze}= t_{\rm GHZ}^{(1)} + 2\pi$. Noise acts only during the post-quench storage stage. (a) GHZ fidelity $F(t)$. (b) GHZ coherence $C(t)$. (c) Sum of populations $p_0(t)$ and $p_1(t)$ of $\ket{0\cdots 0}$ and $\ket{1\cdots 1}$. (d) Quantum Fisher information $\mathcal F_Q^{(y)}(t)/L^2$ with respect to $\hat S_y$. (e) Entanglement depth $\mathcal{D}(t)/L$ inferred from QFI-based $k$-producibility bounds. See text for definitions of $F(t)$, $C(t)$, $p_{0,1}(t)$, $\mathcal F_Q^{(y)}$, and the QFI-based entanglement depth $\mathcal D(t)$.
}
  \label{fig:fig_8_noise_ghz}
\end{figure*}

Figure~\ref{fig:fig_8_noise_ghz} characterizes the stability of the stored GHZ state under relaxation and dephasing. We denote by $\rho(t)$ the many-body density matrix of the system at time $t$, obtained from the Lindblad evolution. As in Sec.~\ref{sec:oat_dicke}, we work in the rotated frame defined by $\hat R_x(-\pi/2)=e^{+{\rm i}(\pi/2)\hat S_x}$ and $\tilde\rho(t)=\hat R_x(+\pi/2)\rho(t)\hat R_x(-\pi/2)$, so that the equatorial GHZ produced by the OAT dynamics is mapped to the computational GHZ state within subspace spanned by $\{\ket{0\cdots 0},\ket{1\cdots 1}\}$.

The memory performance is first quantified by the GHZ fidelity in Fig.~\ref{fig:fig_8_noise_ghz}(a)
\begin{equation*}
F(t)=\operatorname{Tr}\!\left[\ket{{\rm GHZ}_\phi}\bra{{\rm GHZ}_\phi}\,\tilde\rho(t)\right]
=\bra{{\rm GHZ}_\phi}\tilde\rho(t)\ket{{\rm GHZ}_\phi},\qquad
\ket{{\rm GHZ}_\phi}=\frac{\ket{0\cdots 0}+e^{{\rm i}\phi}\ket{1\cdots 1}}{\sqrt{2}} \ ,
\end{equation*}
with the phase $\phi$ fixed by the unitary evolution at $t=t_{\rm GHZ}^{(1)}$. To separate coherence loss from population leakage, we monitor the edge populations
\[
p_0(t)=\bra{0\cdots 0}\tilde\rho(t)\ket{0\cdots 0},\qquad
p_1(t)=\bra{1\cdots 1}\tilde\rho(t)\ket{1\cdots 1},
\]
and the off-diagonal coherence between these two components in Fig.~\ref{fig:fig_8_noise_ghz}(b)
\[
C(t)=2\left|\bra{0\cdots 0}\tilde\rho(t)\ket{1\cdots 1}\right|.
\]
In Fig.~\ref{fig:fig_8_noise_ghz}(c), we examine $p_0(t)+p_1(t)$ to quantify the population remaining in the GHZ manifold.

\tcr{To assess metrological usefulness, we compute the quantum Fisher information \cite{QFI_Braunstein}
\[
\mathcal F_Q[\rho(t),\hat A]
=2\sum_{k,\ell}\frac{(\lambda_k-\lambda_\ell)^2}{\lambda_k+\lambda_\ell}
\left|\matrixel{k}{\hat A}{\ell}\right|^2,
\]
with $\rho(t)=\sum_k\lambda_k\ket{k}\bra{k}$ and $\hat A=\hat S_y$. In Fig.~\ref{fig:fig_8_noise_ghz}(d) we plot $\mathcal F_Q^{(y)}(t)/L^2$, so that an ideal GHZ corresponds to a value of order unity, reflecting Heisenberg scaling $\mathcal F_Q\sim L^2$ \cite{QFI_heisen_2, QFI_heisen_1}.}

\tcr{Finally, we quantify genuine multipartite entanglement through the entanglement depth $\mathcal{D}$. 
By definition, $\mathcal{D}$ \cite{Sorensen2001} is the size of the largest subset of qubits that must be genuinely entangled with each other: $\mathcal{D}=1$ corresponds to fully separable states, while $\mathcal{D}=L$ implies that all $L$ qubits share genuine multipartite entanglement. Entanglement depth captures globally shared entanglement in contrast to merely pairwise entanglement, making it particularly meaningful for states such as GHZ, whose defining feature is macroscopic, system-wide entanglement.}

\tcr{Operationally, we infer a "minimum" depth from the maximal QFI over collective directions, 
$\mathcal F_Q^{\rm max}(t)=\max\{\mathcal F_Q[\rho(t),\hat S_x],\mathcal F_Q[\rho(t),\hat S_y],\mathcal F_Q[\rho(t),\hat S_z]\}$. 
For an $L$ qubit $k$-producible state one has the bound $\mathcal F_Q^{\rm max}\le s k^2+r^2$, with 
$s=\lfloor L/k\rfloor$ and $r=L-sk$~\cite{QFI_heisen_2,Toth2012}. 
A violation of this inequality \emph{certifies that the entanglement depth is at least $k+1$}. The resulting $\mathcal{D}(t)$, normalized by $L$, is shown in Fig.~\ref{fig:fig_8_noise_ghz}(e).}

\tcr{Several trends are apparent. The GHZ fidelity in Fig.~\ref{fig:fig_8_noise_ghz}(a) remains comparatively robust over the simulated storage window. The QFI in Fig.~\ref{fig:fig_8_noise_ghz}(d) decays faster than the fidelity, reflecting its stronger sensitivity to small losses of coherence, yet it remains sizable for a substantial fraction of the post-quench evolution. Figures~\ref{fig:fig_8_noise_ghz}(b) and~\ref{fig:fig_8_noise_ghz}(c) show that the coherence $C(t)$ decreases more rapidly than the total GHZ weight $p_0(t)+p_1(t)$, consistent with dephasing suppressing the off-diagonal element between the two macroscopic branches before significant population leakage occurs. Finally in Fig.~\ref{fig:fig_8_noise_ghz}(e), the entanglement depth remains maximal, $\mathcal{D}(t)=L$, over the simulated times for smaller system sizes, while a reduction of the certified depth becomes visible only for larger systems (here starting at $L=12$), reflecting the increased sensitivity of larger GHZ states to decoherence. Overall, for representative superconducting coherence times, the emergent-Hamiltonian protocol maintains a fairly high fidelity with the target GHZ state over the simulated durations. While metrological gain degrades earlier than fidelity, the QFI-based witness continues to certify large, and often maximal, multipartite entanglement depth over a substantial portion of the storage window.}

\end{document}